\documentclass[fleqn,usenatbib]{mnras}
\usepackage{newtxtext,newtxmath}

\usepackage[T1]{fontenc}
\usepackage{ae,aecompl}

\usepackage{graphicx}	
\usepackage{amsmath}	
\usepackage{amssymb}	
\DeclareMathOperator{\sech}{sech}

\title[Simulating star formation in M51]{Simulations of the star-forming molecular gas in an interacting M51-like galaxy}

\author[R. G. Tress et al.]{Robin G. Tre{\ss},$^{1}$\thanks{E-mail: robin.tress@uni-heidelberg.de}
Rowan J. Smith,$^{2}$
Mattia C. Sormani,$^{1}$
Simon C. O. Glover,$^{1}$
\newauthor
Ralf S. Klessen, $^{1,3}$
Mordecai-Mark Mac Low$^{4,5}$
and Paul C. Clark$^{6}$
\\
$^1$Universit\"{a}t Heidelberg, Zentrum f\"{u}r Astronomie, Institut f\"{u}r Theoretische Astrophysik, Albert-Ueberle-Str. 2, 69120 Heidelberg, Germany \\
$^2$Jodrell Bank Centre for Astrophysics, School of Physics and Astronomy, University of Manchester, Oxford Road, Manchester M13 9PL, UK\\
$^3$Universit\"at Heidelberg, Interdisziplin\"ares Zentrum f\"ur Wissenschaftliches Rechnen, Im Neuenheimer Feld 205, 69120 Heidelberg, Germany \\
$^4$Department of Astrophysics, American Museum of Natural History, New York, NY 10024, USA\\
$^5$Center for Computational Astrophysics, Flatiron Institute, New York, NY, 10010, USA \\
$^6$School of Physics and Astronomy, Queen's Buildings, The Parade, Cardiff University, Cardiff, CF24 3AA \\
}

\date{Accepted XXX. Received YYY; in original form ZZZ}

\pubyear{2019}

\begin{document}
\label{firstpage}
\pagerange{\pageref{firstpage}--\pageref{lastpage}}
\maketitle

\begin{abstract}
We present here the first of a series of papers aimed at better understanding the evolution and properties of giant molecular clouds (GMCs) in a galactic context. We perform high resolution, three-dimensional {\sc arepo} simulations of an interacting galaxy inspired by the well-observed M51 galaxy. Our fiducial simulations include a non-equilibrium, time-dependent, chemical network that follows the evolution of atomic and molecular hydrogen as well as carbon and oxygen self-consistently. Our calculations also treat gas self-gravity and subsequent star formation (described by sink particles), and coupled supernova feedback. In the densest parts of the simulated interstellar medium (ISM) we reach sub-parsec resolution, granting us the ability to resolve individual GMCs and their formation and destruction self-consistently throughout the galaxy. In this initial work we focus on the general properties of the ISM with a particular focus on the cold star-forming gas. We discuss the role of the interaction with the companion galaxy in generating cold molecular gas and controlling stellar birth. We find that while the interaction drives large-scale gas flows and induces spiral arms in the galaxy, it is of secondary importance in determining gas fractions in the different ISM phases and the overall star-formation rate. The behaviour of the gas on small GMC scales instead is mostly controlled by the self-regulating property of the ISM driven by coupled feedback. 
\end{abstract}




\section{Introduction}

In the cold and dense molecular phase of the interstellar medium (ISM), it is much easier to trigger runaway gravitational collapse, which makes giant molecular clouds (GMCs) the preferred birth-site of stars. Key questions in the study of star formation (SF) on galactic scales therefore include how the gas gets to these densities and temperatures, and what controls the amount of cold gas with respect to the other thermal phases of the ISM. A thorough understanding of the properties, evolution and dynamics of the ISM and especially of the cold molecular phase, is a vital step towards a predictive theory of SF \citep{MacLow&Klessen2004, McKee&Ostriker2007, Klessen&Glover2016}.

The ISM is composed of three main thermal phases: a hot ($T \sim 10^{6} \: {\rm K}$) ionised phase produced by mechanical energy input from supernovae and stellar winds; a warm ($T \sim 10^{4} \: {\rm K}$) phase that can be further subdivided into largely ionised gas (found e.g.\ in H~{\sc ii} regions around massive stars or the diffuse ionised medium) or largely neutral atomic gas; and a cold phase composed of a mix of atomic and molecular gas \citep{McKee1977}. Although the warm atomic phase is generally close to thermal equilibrium \citep[e.g.][]{Wolfire1995}, the ISM is a rich and dynamic system and perturbations can generate thermal instabilities that lead to runaway cooling and the formation of cold, dense gas clouds. Clouds that are dense and massive enough to shield themselves from the interstellar radiation field develop high molecular fractions, becoming GMCs. General causes for these perturbations include gravitational instabilities, cloud-cloud collisions of warm atomic gas, and shocks in the turbulent ISM \citep[e.g.][and references therein]{Klessen&Hennebelle2010, Smith+2014, Dobbs2014}. Once formed, GMCs may be disrupted by feedback from the stars forming within them \citep{Krumholz+2014} or by external processes such as galactic shear \citep[e.g.][]{Jeffreson+2018}.

Larger-scale processes can play a fundamental role as well: spiral arms and bars can gather together warm gas, triggering the formation of cold clouds and initiating the SF process. The global rotation curve of a galaxy is a central parameter, directly affecting the Toomre Q parameter and thus the local stability against gravitational collapse of the disc \citep{Li+2005} as well as controlling the local shear experienced by the ISM. 

Most noticeably, in the question of what controls SF, galaxy interactions and mergers play a prominent role. Mergers are often associated with bursts of SF \citep{Larson&Tinsley1978, Lonsdale+1984, Barton+2000, Ellison+2008, Renaud+2014} and the most vigorously star-forming galaxies can all be morphologically interpreted as merging pairs \citep{SandersAndMirabel1996}. These enhancements in SF may be triggered by the collapse of previously stable gas due to cloud collisions and gas compression in tidally-induced spiral arms \citep{Toomre&Toomre1972}. In addition, mergers can result in a  substantial redistribution of the angular momentum of the gas and also in the formation of bars \citep{Mihos&Hernquist1996}, both of which act to drive large-scale gas flows towards the centre of the most massive galaxy, leading to a nuclear starburst. 

Simulations \citep[e.g.][]{Cox+2006, DiMatteo+2007, Renaud+2014} show that the SF histories of these interactions exhibit an increase in SF immediately after the pericentric passage of the companion and then again during the coalescence phase with an increase of the SF rate (SFR) of at least a factor of two with respect to the isolated case. 

However, the details are case-specific and initial conditions such as orbital parameters, mass ratios, gas fractions and the initial stability of the isolated disc all play an important role \citep{DiMatteo+2007, Cox+2008} such that not every merger is immediately linked to an enhanced SFR. \citet{Cox+2008} showed that there is a strong correlation with the mass ratio and found that for high mass differences between the two galaxies it is questionable whether the interaction drives any SF whatsoever, since the tidal disturbance is small. Even for similar masses the merger can, for example, remove a large amount of gas from the galaxy during the first encounter via tidal tails. If this gas cannot fully re-accrete during the coalescence phase, the galaxy will be unable to form significantly more stars than in its isolated state \citep{DiMatteo+2007}.

In addition, if the isolated galactic disc is mainly Toomre unstable it will already be collapsing and radially flowing towards the centre. In this case, the disc is maximally star forming and the rate is self-regulated by the energy input of stellar feedback to counterbalance the midplane pressure exerted by the disc \citep{Ostriker&Shetty2011, Shetty&Ostriker2012}, combined with the energy input from radial infall \citep{Krumholz+2018}. If the interaction cannot significantly increase the midplane pressure and drive radial inflow, it is unlikely that the SFR will be enhanced.

In recent decades, numerical simulations of the ISM have improved substantially, reaching ever higher resolutions and including more and more physical ingredients. There has been important progress in understanding the relative importance of different physical processes and their direct effect on the ISM phases and thus on the regulation of SF \citep[e.g.][]{Dale+2014, Gatto+2015, Walch+2015, Kim&Ostriker2017, Peters+2017, Hill+2018,Rahner+2019}. However simulations that can resolve scales smaller than entire GMCs rarely include larger galactic scale phenomena\footnote{Exceptions are dwarf galaxy simulations where the total gas mass is small and more detailed studies are possible \citep[e.g.][]{Hu+2016, Hu+2017, Emerick+2019}.} and are often carried out using highly idealised setups, such as isolated or colliding clouds or kpc-sized portions of the stratified galactic disc.

Larger-scale simulations typically rely on sub-grid models to follow the SF process, which abstract the complex nature of the ISM on the cloud and sub-cloud scales \citep[e.g.][]{Hopkins+2014, Vogelsberger+2014, Shaye+2015, Pillepich+2018}. In these cases, the transition from atomic to molecular gas and the composition of the ISM on the scale of individual GMCs are at best only marginally resolved. In spite of this, these models have proved very successful for developing a general understanding of the dynamics of the ISM in galaxies and its global SF properties. However, their predictive power starts to become questionable on smaller scales and research bridging the gap between these large-scale models and detailed simulations of individual clouds has only recently started to become computationally viable.  

The details of how a galaxy encounter affects its cold molecular gas content and the following SF on the level of individual molecular clouds remains an open question. For instance, it has been debated whether the increase of SF during the encounter reflects an increase in the available molecular gas reservoir or whether it follows from a higher SF efficiency with strong arguments favouring both sides \citep{Cox+2006, Krumholz+2012, Pan+2018}. Some of the observational studies endorsing the higher efficiency scenario assume a different conversion factor between CO and H$_2$ with respect to more quiescent galaxies \citep{Daddi+2010} and some numerical studies seem to hint at such a scenario \citep{Renaud+2019}. But ultimately these highly interesting questions can only be fully addressed with models capable of properly resolving the molecular phase of the ISM.

In this paper, we study the ISM of a galaxy undergoing a merger with a particular focus on the molecular gas. We try to understand how the encounter affects the gas properties with the help of high-resolution galactic-scale simulations carried out using the {\sc arepo} moving-mesh code \citep{Springel2010}. One key goal of our current study is to quantify the relative importance of local feedback and global dynamical processes for regulating the SFR and shaping the molecular phase of the ISM.  

We take inspiration for our model from the well-studied interacting galaxy M51 (also known as NGC 5194 or the Whirlpool galaxy). M51 is a nearby and almost face-on example of a galaxy currently undergoing a merger, in this case with its smaller neighbour NGC 5195, with a mass ratio around one-half \citep{Schweizer1977, Mentuch+2012}. Owing to this interaction, it displays a prominent grand-design spiral pattern. Because of this, plus its relative proximity and favourable inclination, it has been the target of many observational studies, of which the most important for our purposes is probably the PdBI Arcsecond Whirlpool Survey (PAWS; \citealt{Schinnerer+2013}), which mapped CO emission on scales down to $\sim 40$~pc, comparable to the size of individual GMCs.

Our simulations are not the first to attempt to model M51. We follow the lead of \citet{Dobbs+2010}, who simulated the gas and the stars of a system inspired by the present-day M51 system. However, they used an isothermal equation of state for the gas, preventing them from studying the cold gas distribution or the properties of individual GMCs. More recently, \citet{Pettitt+2017} performed smoothed particle hydrodynamics (SPH) simulations of an interacting galaxy morphologically similar to M51 using a more sophisticated thermal treatment. However, the SPH particle mass in their simulations was $2000 \: {\rm M_{\odot}}$, rendering them unable to resolve all but the largest GMCs. The simulations presented here have considerably higher resolution, down to a few M$_\odot$, allowing us to resolve a much broader range of GMCs.

Our paper is organised as follows. In Section \ref{sec:Methods} we describe our model, the numerical methods and the initial conditions of our simulation. We describe the outcome of our simulations in Section \ref{sec:Results} with a particular emphasis on the ISM properties and the SFR. We then proceed to analyse the role of the galactic interaction on the ISM phases in Section \ref{sec:Discussion} by comparing simulations of an isolated and an interacting galaxy with the same initial properties. We discuss the limitations and problems of our model in Section \ref{sec:caveats} and summarise our findings in Section \ref{sec:Conclusions}.

\section{Methods}
\label{sec:Methods}

\subsection{Numerical code}
\label{sec:Arepo} 
Our simulations were performed with {\sc arepo} \citep{Springel2010}, which is a moving-mesh hydrodynamic code coupled with an $N$-body gravity solver. 

The fundamental conservation laws needed to describe the evolution of an unmagnetised\footnote{Although the ISM of M51 is known to be magnetised \citep[see e.g.][]{Fletcher+2011}, we restrict our attention here to the unmagnetised case for simplicity, and defer any investigation of MHD effects to a future study.} fluid are conservation of mass, momentum and energy: 
\begin{equation}
    \label{eq:Continuity}
    \frac{\partial \rho}{\partial t} + \nabla \cdot (\rho \boldsymbol{u}) = 0;
\end{equation}
\begin{equation}
    \label{eq:euler}
    \frac{\partial \boldsymbol{u}}{\partial t} + \boldsymbol{u} \cdot \nabla \boldsymbol{u} = - \frac{\nabla P}{\rho} - \nabla \Phi;
\end{equation}
\begin{equation}
    \label{eq:energy}
    \frac{\partial \rho e}{\partial t} + \nabla \cdot [(\rho e + P)\boldsymbol{u}] = \rho \Dot{Q} + \rho \frac{\partial \Phi}{\partial t}.
\end{equation}
Here $\rho$ is the mass density, $\boldsymbol{u}$ is the velocity field, $P$ is the thermal pressure, $\Phi$ is the gravitational potential, $e = e_{\rm th} + \Phi + \boldsymbol{u}^2 / 2$ is the total energy per unit mass, and $e_{\rm th}$ is the thermal energy per unit mass. The term $\Dot{Q}$ hides all the complexity of the chemical and radiative cooling and heating processes described in Section~\ref{sec:SGChem} below.

To close the system, we use an ideal gas equation of state,
\begin{equation}
    \label{eq:eos}
    P = (\gamma - 1) \rho e_{\rm th},
\end{equation}
where $\gamma$ is the adiabatic index. We set $\gamma = 5/3$ throughout the simulation, even in gas which is predominantly molecular. We justify our neglect of the rotational degrees of freedom of the molecular gas by noting that in our simulations, gas with a high molecular fraction is typically too cold to excite the rotational energy levels of H$_{2}$.

These fluid equations are then solved in 3D on a time-dependent mesh constructed by computing the Voronoi tesselation of the domain given a set of mesh-generating points. By assigning to each mesh-generating point the local velocity of the fluid, the grid can naturally follow the flow and continuously adapt the configuration of the cells which will approximately retain constant mass. As a quasi-Lagrangian scheme, the resolution of {\sc arepo} strongly depends on the fluid density. Moreover, instead of inferring the necessary time-step globally based on the Courant criterion, this is done locally and every cell is evolved in time based on its local conditions. The code is therefore able to efficiently deal with problems having a large dynamical range both in space and in time, which is necessary to study a multi-scale problem such as the ISM dynamics of a galaxy.

Other strengths of the {\sc arepo} code include its (nearly) Galilean invariance, its good shock treatment, its minimization of advection errors and the lack of an underlying preferential mesh geometry. At every interface between cells, the code finds the flux by solving the Riemann problem in the rest-frame of the interface. Since the cells are moving approximately at the local fluid velocity, these fluxes are kept minimal and advection errors are thus small. Furthermore in this way the solution is independent from the chosen frame of reference and best suited to study problems where there is no preferred flow direction.

To compute the gravitational potential, {\sc arepo} uses a tree-based approach adapted from an improved version of {\sc gadget-2} \citep{Springel2005}. The contribution of the gas cells is included by treating each as a point mass located at the centre of the cell, with an associated gravitational softening. This softening changes as the gas cell grows or shrinks, with a lower limit in our simulations of $0.1$~pc.

The self-gravity of the gas is mainly important on small scales when local gravitational runaway collapse sets in, leading to SF. On larger scales, the gravity is dominated by the dark matter and the old stellar population of the galaxy. We follow the behaviour of both components by simulating the dynamical evolution of a set of representative dark matter (DM) and star particles, which are assumed to be collisionless, meaning that they only enter into the gravity calculation. The masses and softening lengths chosen for these particles are discussed in Section~\ref{sec:resolution} below.

\subsection{Chemical network}
\label{sec:SGChem}
The chemical evolution of the gas is modelled using the NL97 network of \citet{Glover+2012}, which combines the network for hydrogen chemistry presented in \citet{Glover+2007a,Glover+2007b} with a highly simplified treatment of CO formation and destruction developed by \citet{Nelson+Langer1997}. The NL97 network was first implemented in {\sc arepo} by \citet{Smith+2014} and has subsequently been used in a number of studies with this code \citep[e.g.][]{Bertram+2015,Sormani+2018}. The effects of dust shielding and H$_{2}$ and CO self-shielding from the non-ionizing UV part of the interstellar radiation field are modelled using the {\sc TreeCol} algorithm developed by \citet{Clark+2012}. The background radiation is taken to be spatially constant at a solar-neighbourhood value \citep{Draine1978}.

We also solve for the thermal evolution of the gas due to radiative heating and cooling, which we compute simultaneously with the chemical evolution. We use a detailed atomic and molecular cooling function, the latest version of which is described in \citet{Clark+2018}. Of particular note here is our treatment of the cooling of the gas at high temperatures ($T \gg 10^{4}$~K) owing to the effects of permitted atomic transitions, since this is of great importance for modelling the effects of supernova feedback. We treat cooling due to transitions in atomic hydrogen in a fully non-equilibrium fashion using the H and electron abundances computed in our chemical model. For cooling due to transitions in He and metals we use the collisional ionization equilibrium cooling rates tabulated in \citet{Gnat+Ferland2012} instead. 

We finally impose a temperature floor of $20$~K on the simulated ISM. Without this floor, the code can occasionally produce anomalously low temperatures in cells close to the resolution limit undergoing strong adiabatic cooling, with unfortunate effects on the stability of the code. Since the equilibrium gas temperature is comparable to or larger than 20~K throughout the full range of densities resolved in our simulation \citep[see e.g.\ the temperature-density plots in the high resolution simulations of individual clouds presented in][]{Clark+2018}, we do not expect the presence of this temperature floor to have any effects on the results of our simulations.

\subsection{Sink particles}
\label{sec:SinkParticles}
\begin{figure}
	\includegraphics[width=\columnwidth]{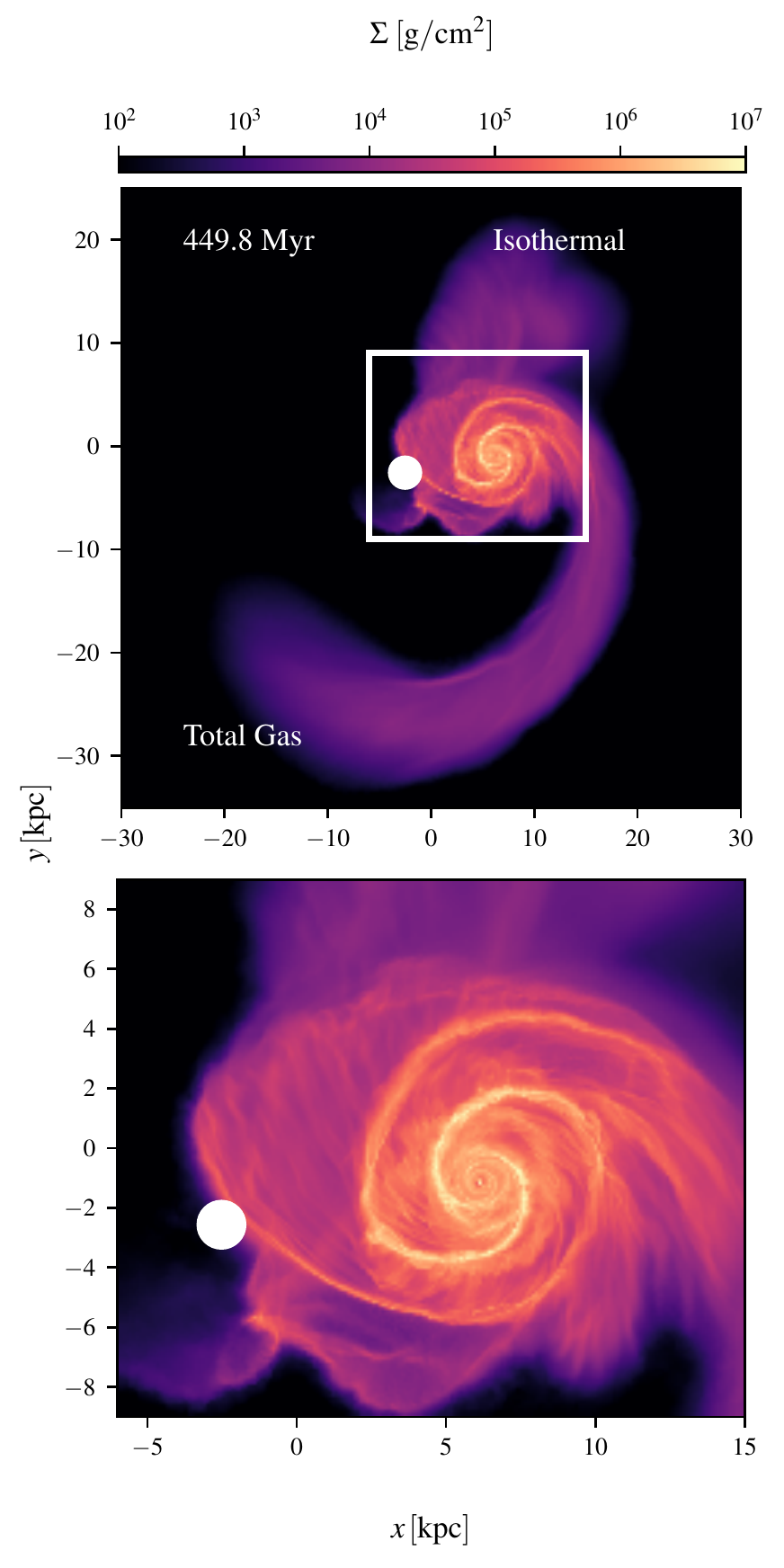}
    \caption{Total gas column density map of the isothermal test run at $t \simeq 450$~Myr.}
    \label{fig:rhoproj_isothermal}
\end{figure}
In order to form stars, gas must undergo gravitational collapse and increase its density by many orders of magnitude from typical GMC densities to protostellar densities. Despite {\sc arepo}'s adaptive capabilities, it is not computationally feasible to simulate the whole of this process in simulations of the scale presented here. We therefore adopt a technique widely used in computational studies of SF and replace the densest gravitationally-bound collapsing regions in the simulation with collisionless sink particles. 

Sink particles, hereafter referred to as sinks, are mainly used in high resolution simulations of individual clouds where the SF process can be spatially and temporally resolved reasonably well. Typically galactic-scale simulations cannot resolve the collapse within GMCs, and tend to avoid using accreting sink particles to represent SF. Instead non-accreting star particles are often employed where particles of a given mass representing stars are stochastically created in the densest parts of the ISM \citep[e.g.][]{Katz1992, Katz+1996, Stinson+2006}. These schemes are often fine-tuned to reproduce the Schmidt-Kennicutt relation \citep{Schmidt1959, Kennicutt1989, Kennicutt1998}; although they tend to produce a healthy matter cycle in the ISM, their power to predict SF is strongly limited. Since our resolution reaches below GMC scales, we use accreting sink particles that we describe here and discuss possible caveats and limitations in Section \ref{sec:caveats}. 

Following \citet{Bate+1995} and \citet{Federrath+2010}, sink particles form if within an accretion radius $r_{\rm acc}$, a region above a density threshold $\rho_{\rm c}$ satisfies these criteria:
\begin{enumerate}
\item The gas flow in this region is converging. To establish this, we require not only that the velocity divergence is negative ($\nabla \cdot v < 0$) but also that the divergence of the acceleration is negative ($\nabla \cdot a < 0$).  
\item The region is located at a local minimum of the potential.
\item The region is not situated within the accretion radius of another sink and also will not move within the accretion radius of a sink in a time less than the local free-fall time.
\item The region is gravitationally bound, i.e. $U > 2 (E_{\rm k} + E_{\rm th})$, where $U = G M^2 / r_{\rm acc}$ is the gravitational energy of the region within the accretion radius, $E_{\rm k} = 1/2 \sum_i m_i \Delta v_i^{2}$ is the total kinetic energy of all gas particles within the accretion radius with respect to the centre of collapse, and $E_{\rm th} = \sum_i m_i e_{{\rm th}, i}$ is the total internal energy of the same region.
\end{enumerate}
These criteria help to ensure that a region of the gas is only converted into a sink particle if it is truly self-gravitating and collapsing. 

If a gas cell satisfies all of the above criteria it is turned into a collisionless sink particle. These are then allowed to accrete mass during the simulation. If a gas cell within $r_{\rm acc}$ is denser than the threshold density and the gas is also gravitationally bound to the sink particle, then we move an amount of mass
\begin{equation}
\Delta m = \left(\rho_{\rm cell} - \rho_{\rm c}\right) V_{\rm cell}
\end{equation}
from the cell to the sink, where $\rho_{\rm cell}$ is the initial gas density in the cell and $V_{\rm cell}$ is its volume. Afterwards the new density of the cell is simply the threshold density $\rho_{\rm c}$. We also update appropriately any other quantities in the cell that depend on the mass, such as the total momentum or kinetic energy. In the case where a given gas cell is located within the accretion radii of multiple sink particles, we place the accreted mass from it onto the sink to which the gas is most strongly bound.

\begin{table}
	\centering
	\begin{tabular}{lcr} 
		\hline
		$\rho_{\rm c}$ (g~cm$^{-3}$)		 & $10^{-21}$ \\
		$r_{\rm acc}$ (pc)       & $2.5$ \\
        Softening length (pc)    & $2.5$ \\
        \hline
        $\epsilon_{\rm SF}$      & $0.05$ \\
        $r_{\rm sc}$ (pc)   & $5.0$ \\
          
		\hline
	\end{tabular}
        \caption{Parameters of the sink particles; $\rho_{\rm c}$ is the sink density threshold, $r_{\rm acc}$ is the accretion radius, $\epsilon_{\rm SF}$ is the SF efficiency, and $r_{\rm sc}$ is the scatter radius of SNe around the sink. For details see the text.\label{tab:sinks}}
\end{table}

As discussed in more detail in \citet{Springel2010}, {\sc arepo} makes use of a hierarchical time-stepping scheme in which individual gas cells and collisionless particles can have different timesteps, meaning that at any given time not every cell and particle is active. Accretion onto a sink particle is allowed only when the sink particle itself and the candidate gas cells are active. In order to ensure that we do not miss accretion from cells that spend only a short time within the sink accretion radius, we make sure that the timestep used to evolve the sinks is the same as the shortest one of the gas cells. 

In order to properly follow the hierarchical collapse and correctly resolve the underlying fragmentation, we ensure that the local Jeans length is resolved by at least four resolution elements (\citealt{Truelove+1997}; see also \citealt{Federrath+2011} for further discussion). In the densest and coldest parts of GMCs, the Jeans length can, however, become prohibitively small. We therefore choose to stop refining for densities above $\rho_{\rm lim} = 10^{-21}$~g~cm$^{-3}$ and we set $\rho_c = \rho_{\rm lim}$. This is a good compromise between excellent resolution in the collapsing regions, and computational performance. The chosen density threshold is also well above the critical density for H$_2$ formation \citep{Smith+2014} such that this process is fully captured by the simulation. 

The accretion radius is chosen such that at the given threshold density, several cells fall inside $r_{\rm acc}$ given the local size of the cells. The gravitational softening length of the collisionless sink particles is set to the same value as $r_{\rm acc}$, as this ensures that the gravitational potential is not altered much due to the infall of mass onto a sink, while at the same time limiting the size of the gravitational acceleration produced within $r_{\rm acc}$, which otherwise would have a detrimental effect on performance. The main parameters that characterise the sink particles used in our study are listed in Table \ref{tab:sinks}. 

\subsection{Feedback}
\label{sec:Feedback}
To form molecular clouds we need to capture the proper cooling and chemical evolution of the dense ISM. Our chemical network in conjunction with our model of the self-shielding properties of the molecular gas from the ambient dissociating UV radiation is the key to achieving this. 

To capture the disruption of molecular clouds, on the other hand, it is necessary to introduce some feedback mechanism. In principle galactic shear can disperse dense molecular clouds, but in general this is insufficient to produce realistic cloud lifetimes and masses throughout the galactic disc \citep{Jeffreson+2018}. GMCs turn out to be too massive and too long-lived when shear is the only disruptive process. 

Supernovae (SNe) randomly distributed with respect to the gas have also been shown to be ineffective at destroying molecular clouds. On the contrary, they help to pile up gas into dense, compact regions, resulting in massive molecular cloud complexes with extremely long lifetimes \citep{Gatto+2015, Walch+2015}. Therefore, a feedback mechanism coupled to the SF is needed that disrupts the clouds from within. In our present study, we focus on the effects of clustered SNe forming in locations correlated with the sink particles. We have found that this is an effective way to reproduce reasonable lifetimes for our simulated molecular clouds and to generate a healthy matter cycle in the ISM.

Despite our high resolution, we cannot resolve the formation of individual stars in our simulations. Instead, sinks represent small stellar clusters formed during an SF event within a cloud. We relate the mass of stars formed to the mass of the sink by assuming that only a fraction $\epsilon_{\rm SF} = 0.05$ of the mass accreted actually forms stars, since at the scale at which we form the sinks, the SF process is still quite inefficient \citep[see e.g.][]{Evans+2009}. We then attribute a discrete stellar population to the sink based on the method described in \citet{Sormani+2017}. Given an initial mass function (IMF) we populate a set of discrete mass bins using a Poisson distribution with an appropriately chosen mean for each bin. In this way we ensure that the mass distribution of the stars formed in the simulation follows the desired IMF even when the individual sinks are too small to fully sample the IMF. This method also allows us to account for stars formed from mass accreted at later times.

For each star more massive than $8$ M$_\odot$ associated with a sink, we generate an SN event at the end of the lifetime of the star, inferred based on their mass from Table~25.6 of \citet{Maeder2009}. Since the sink represents an entire group of stars that can interact dynamically, we do not assume that the SN occurs exactly at the location of the sink. Instead, we randomly sample the SN location from a Gaussian distribution centred on the particle and with standard deviation $ r_{\rm sc} = 5$~pc.

Since the assumed efficiency of SF within the sink is relatively small, most of the mass in the sink represents gas that should be returned to the ISM once stellar feedback starts. The mass that is not locked up in stars is therefore gradually given back to the ISM with every SN event. Each event ejects a total mass of $M_{\rm ej} = (M_{\rm sink} - M_{\rm stars}) / n_{\rm SN}$, where $M_{\rm sink}$ is the mass of the sink at the time that the supernova occurs, $M_{\rm stars}$ is the mass of stars contained within the sink at that time, and $n_{\rm SN}$ is the remaining number of SN events that the sink harbors. The mass is distributed uniformly within the energy injection region. The temperature of the injection cells is not altered at this stage.

Once the last massive star has reached the end of its lifetime, the sink has a final mass of $M_{\rm stars}$. At this point, we convert it into a collisionless $N$-body particle representing its evolved stellar population. It will then become part of the group of stellar particles that make up the disc and bulge in our simulation. Our base mass resolution for these stellar particles is $\sim 10^4$ M$_\odot$ (see Section \ref{sec:IC}); in order not to lose resolution in computing gravitational interactions, we switch off accretion onto sink particles that have reached this limiting stellar mass content, and allow instead a new sink to form. In this way, sink particles can be seen as maturing star particles and we retain the ability to follow the dynamical evolution of star clusters to some extent. 

Especially in the high resolution simulations, it is not uncommon to have sinks that do not accrete enough mass to form a massive star. In this case we cannot return the gas mass trapped within the sink during an SN event. Instead, after a period of $10$~Myr, if the sink still did not manage to create a massive star, we convert it into a normal star particle and return the remaining mass ($95$\%) to the ISM by uniformly adding it to all gas particles in a surrounding sphere of $R=100$~pc.

In addition to the type II SNe associated with SF, we also account for type Ia SNe, which are produced by the older stellar population in the galaxy. Based on the inferred SF history of M51 \citep{Eufrasio+2017}, we estimate a rate of one SNIa every $250$ years. We create a SN event at this rate\footnote{SNIa actually follow an exponential distribution in time having the given rate as an average.} at the position of a randomly selected star (not sink) particle of the stellar disc and bulge.

To model the supernova energy injection we use a highly modified version of the algorithm first implemented in {\sc arepo} by \citet{Bubel2015}. For every SN event, we calculate the radius of a supernova remnant at the end of its Sedov-Taylor phase based on an assumed SN energy of $10^{51}$~erg and the local mean density $\bar{n}$, which for solar metallicity yields \citep{Blondin1998}
\begin{equation}    
R_{\rm ST} = 19.1 \left(\frac{\bar{n}}{1\mbox{ cm}^{-3}}\right)^{-7/17} \: {\rm pc},
\end{equation}
where in our case $\bar{n}$ is calculated including the contributions from both the ambient gas and also the mass loading of the SN event. We compare this with the radius of the injection region, $R_{\rm inj}$, defined as the size of the smallest sphere around the explosion site that contains 40 grid cells. If $R_{\rm ST} > R_{\rm inj}$, we inject $E_{\rm SN} = 10^{51}$~erg into the injection region in the form of thermal energy and fully ionise the contained gas. If, on the other hand, the Sedov-Taylor phase of the SN remnant is unresolved, then this is a sign that the local density is too high for thermal injection to be reliable. 
If we were to inject thermal energy in this case, then it would be radiated away too quickly, making it unable to generate a strong shock and deposit the correct amount of kinetic energy into the ISM. This is a numerical effect that can be prevented by directly injecting the correct terminal momentum instead. This has been estimated to be \citep[see e.g.][]{Martizzi+2015, Gatto+2015, Kim&Ostriker2015}
\begin{equation}
p_{\rm fin} = 2.6 \times 10^5 n^{-2/17} \; \mbox{ M}_\odot \mbox{ km s}^{-1},
\end{equation}
for a SN of energy $E_{\rm SN} = 10^{51}$~erg and solar metallicity. We do not change the temperature or the ionization state of the region in this case as this would throw off-balance the energy budget in large unresolved regions.

Momentum injection alone can not produce a hot phase in the ISM. By keeping the injection radius small we minimise the number of occasions on which we must inject momentum rather than thermal energy. On the other hand, taking too small an injection radius leads to unphysically anisotropic momentum injection. We have found through experimentation that defining $R_{\rm inj}$ such that a total of $40$ grid cells are contained within a sphere of that radius seems to offer the best trade-off between minimizing the number of momentum injection events and minimizing the impact of grid noise and anisotropic expansion on the evolution of the individual remnants. We note that this mixed approach of injecting thermal energy in regions where $R_{\rm ST}$ is resolved and momentum in regions where this is not the case is not new. Similar methods have been successfully used by a number of other authors to study the impact of SN feedback on the ISM \citep[see e.g.][]{Kimm+2014,Hopkins+2014,Walch+2015,Simpson+2015,Kim&Ostriker2017}.

Finally, we note that SNe are not the only type of feedback associated with SF. For example, stellar winds and radiation from young stars also play an important role in dispersing GMCs, particularly since they act much earlier than SN feedback \citep[e.g.][]{Dale+2014,Inutsuka+2015,Gatto+2017,Rahner+2018, Rahner+2019}. However, it remains computationally challenging to include all of these forms of feedback in simulations with the scale and resolution of those presented here. Therefore, in our initial study we restrict our attention to the effects of SN feedback and defer an investigation of other feedback processes to future work.

\subsection{Initial conditions}
\label{sec:IC}
\begin{figure}
	\includegraphics[width=\columnwidth]{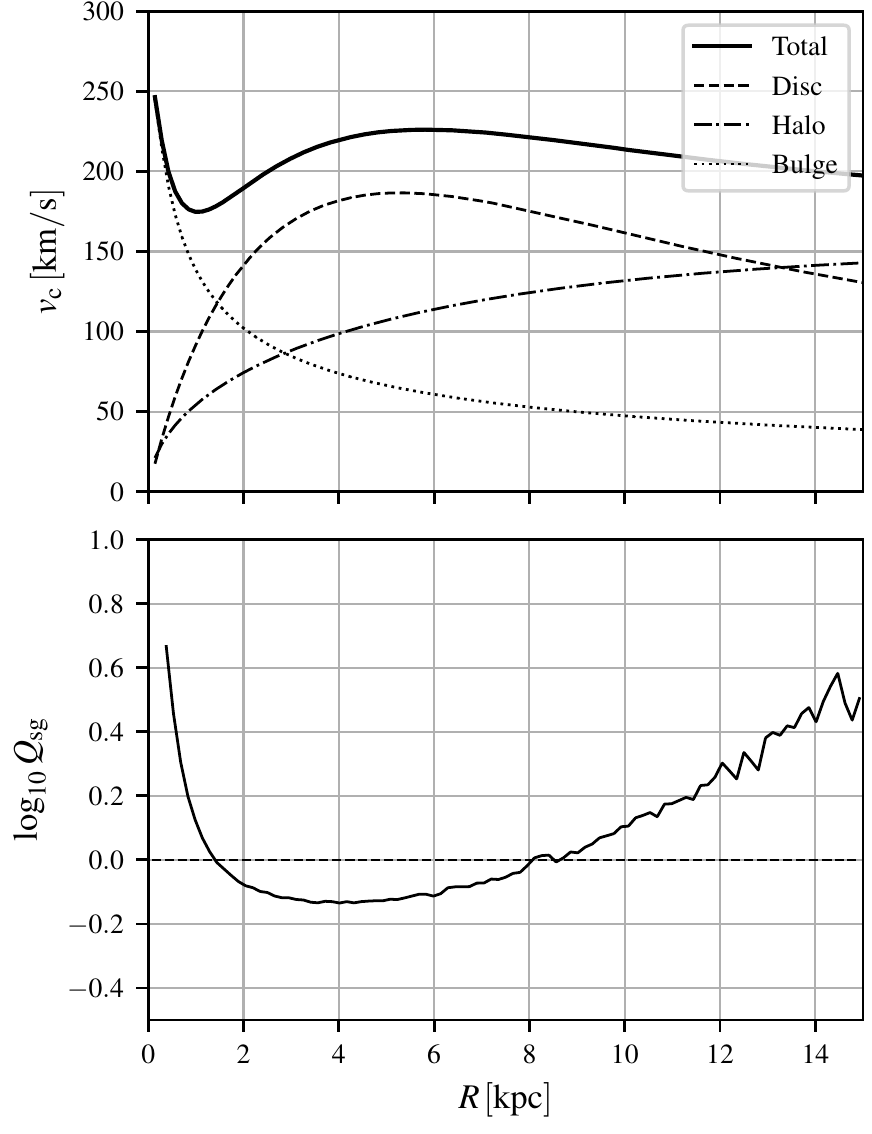}
    \caption{{\em Top}: initial circular velocity curve for the galaxy model. The solid line shows the behaviour for the full galaxy, while the other lines show the contribution of the individual components. {\em Bottom}: combined star-gas Toomre parameter using the equation derived by \citet{Rafikov2001} for the chosen initial conditions. Everything below the dashed line is Toomre-unstable.}
    \label{fig:vc}
\end{figure}

\begin{table}
	\centering
	\caption{Parameters of the different galaxy components.
	\label{tab:mass}}
	\begin{tabular}{lccr} 
		\hline
		 & Mass (M$_\odot$) & Scale length (kpc) & $h_z$ (kpc)\\
		\hline
		DM Halo 	 & $6.04 \times 10^{11}$ & 28.7 & -\\
		Bulge   	 & $5.30 \times 10^9    $ & $9.03 \times 10^{-2} $& -\\
		Stellar disc & $4.77 \times 10^{10} $ & 2.26 & $0.3$\\
		Gas disc 	 & $5.30 \times 10^9    $ & 2.26 & $0.3$\\
		\hline
	\end{tabular}
\end{table}
 
\begin{table}
	\centering
	\caption{Initial conditions of the companion galaxy. }
	\label{tab:NGC5195}
	\begin{tabular}{lclc}
          \hline
          position              & value (kpc)    & velocity         & value (km~s$^{-1}$)   \\
        \hline
		$x_0$      & $-21.91$	&	$v_{x_0}$ 	& $73.2$  \\
		$y_0$ 	   & $-8.44$	&	$v_{y_0}$ 	& $-31.2$ \\
		$z_0$ 	   & $-4.25$	&	$v_{z_0}$ 	& $188.6$ \\
		\hline
	\end{tabular}
\end{table}

We set up a disc galaxy in isolation that consists of four different components: a dark matter halo, a stellar bulge, a stellar disc and a gaseous disc. 

The bulge and the halo follow a spheroidal \citet{Hernquist1990} profile
\begin{equation}
	\rho_{\rm spheroid}(r) = \frac{M_{\rm spheroid}}{2 \pi} \frac{a}{r(r+a)^3}, 
   \label{eq:BulgeProfile}
\end{equation}
where $r$ is the spherical radius, $a$ is the scale-length of the spheroid (the bulge or the halo, depending on which component is considered), and $M_{\rm spheroid}$ is its mass. 

The stellar and gas disc follow a double exponential density profile
\begin{equation}
	\rho_{\rm disc}(R,z) = \frac{M_{\rm disc}}{4 \pi h_z h_R^2} \sech^2 \left(\frac{z}{2 h_z} \right) \exp \left(-\frac{R}{h_R} \right),
   \label{eq:discProfile}
\end{equation}
where $R$ and $z$ are the cylindrical radius and height, and $h_z$ and $h_R$ are the scale-height and scale-length of the disc, respectively. 

We generate the initial conditions using the method and software developed by \citet{Springel+2005}, where the choice of the profile and parameters were cosmologically motivated. The code chooses positions and velocities for the collision-less DM and stellar particles such that the desired equilibrium configuration is established. During setup the gas disc is then created by randomly converting stellar disc particles into gas particles until the desired gas mass fraction is reached. The stellar and gas discs therefore initially follow the same density profile. 

We choose parameters of a typical spiral galaxy, summarised in Table \ref{tab:mass}. The total mass in baryonic matter for the modelled galaxy is chosen to be comparable to the observed baryonic mass of M51a (NGC~5194) which is estimated to be $(5.8 \pm 0.1) \times 10^{10}$~M$_\odot$ \citep{Mentuch+2012}. The scale-length of the disc corresponds approximately to the one listed in \citet{Schruba+2011} with the caveat that the interaction might have affected this observed value. We set the mass of the DM halo under the assumption that its spin parameter is directly connected to the scale-length of the disc by the disc-halo mass ratio and considering the cosmological constraints on this parameter \citep{Hernandez+2007}. The choice of the remaining parameters is motivated instead by the desire to produce a velocity curve (Fig.~\ref{fig:vc}) roughly consistent with observations \citep{Sofue1996, Oikawa+Sofue2014} and to suppress the formation of a strong bar in the simulations of the isolated galaxy. We finally settle on a typical gas-disc mass fraction of $10$~\%. At the highest gas resolution this is the borderline value which is still computationally viable, but we have to consider that this is still only about half of the estimated gas mass in M51 \citep{Mentuch+2012}. 

To model the interaction with the companion galaxy, we follow the approach and initial conditions presented in \citet{Dobbs+2010}, hereafter D10. The companion is described as a single collisionless particle with initial position and velocity given in Table \ref{tab:NGC5195} \citep[taken from][ultimately from \citealt{TheisAndSpinneker2003}]{Dobbs+2010}. Since the companion is reduced to a single particle, we assign a very large gravitational softening of $\epsilon = 3$ kpc to it in order to avoid strong two-body close encounters. This is equivalent to setting the density profile of the galaxy equal to a Plummer sphere with its scale-length equal to the softening length. Given the differences in the model of the main galaxy with respect to D10 we could not reproduce their exact same orbit. However the orbit and the morphological behavior were retrieved by lowering the mass of the particle representing the companion galaxy to $4 \times 10^{10}$ M$_\odot$. Observationally the stellar mass of NGC5195, companion of M51, is estimated by photometry to be $(2.5 \pm 0.2) \times 10^{10}$~M$_{\odot}$ \citep{Mentuch+2012} and thus the mass ratio between the two galaxies is about $0.5$ \citep[see also][]{Schweizer1977}. Our ratio is much lower than that, but given that we recover the morphological behavior of the encounter, the massive particle that we included to represent the companion should rather be seen as the core of its DM halo, while the outer parts do not affect much the dynamics of the encounter.

All these initial assessments were done by running low resolution isothermal simulations with a sound speed $c_{\rm s} =$ 10~km~s$^{-1}$. With the final setup we can reproduce the global morphology of the M51 system (see Fig.\ \ref{fig:rhoproj_isothermal}). We obtain the typical two-armed spiral pattern and the relative position of the two galaxies in the plane of the sky. The second passage of the companion induces a large tidal H~{\sc i} tail also seen in observations. Other characteristics that we reproduce are the peculiar kink in the spiral arm pattern towards the companion (see Fig.\ \ref{fig:rhoproj_isothermal} approximately at position $(2.5,-2.5)$~kpc), and the connecting (only in perspective) arm. All these features were also reproduced in the original work of D10 using SPH. 

Finally we set the initial temperature of the gas to $T = 10^4$ K and consider it to be fully atomic. We assume the ISM to be of solar metallicity throughout. The metallicity of the ISM in the real M51 appears to be slightly super-solar with a small radial metallicity gradient \citep{Croxall+2015}, but we do not expect this minor difference to significantly affect our results.

\subsection{Resolution}
\label{sec:resolution}
\begin{figure}
	\includegraphics[width=\columnwidth]{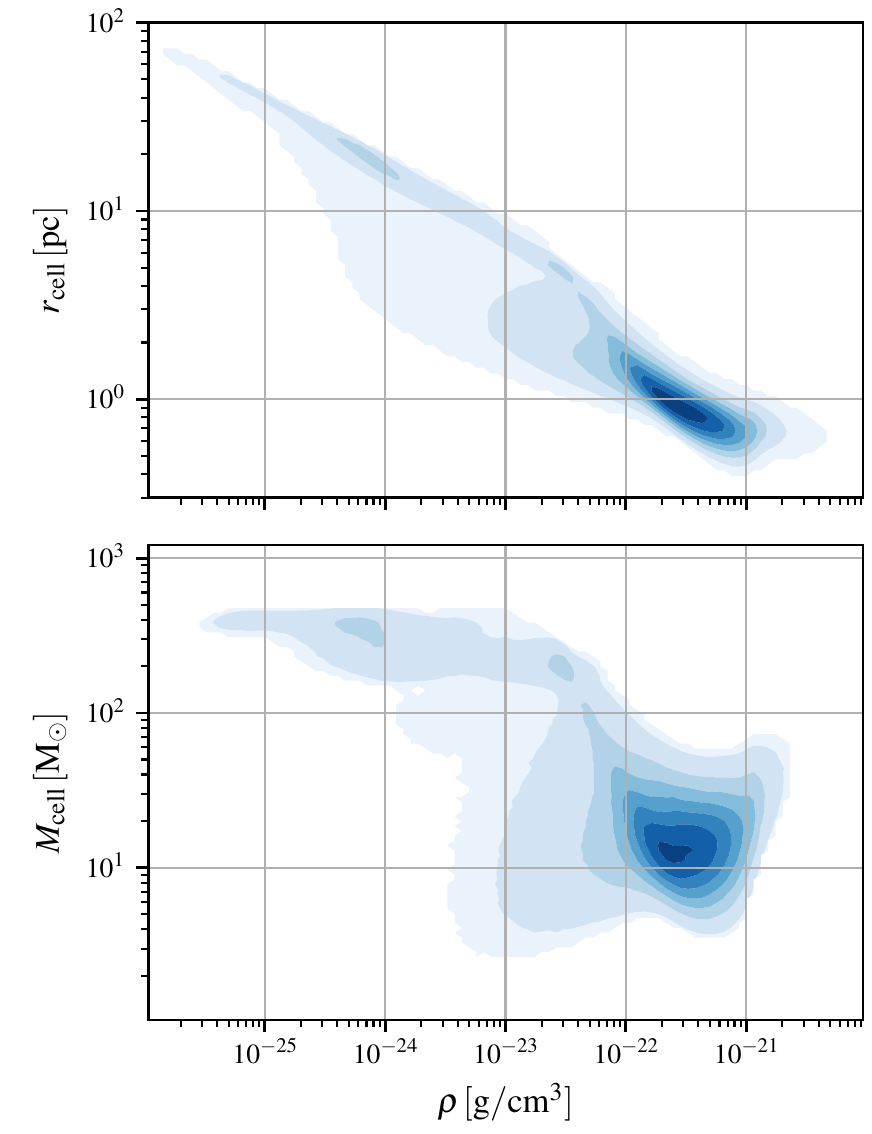}
    \caption{{\em Top}: spatial resolution as a function of local density for the high resolution run. $r_{\rm cell}$ is the radius of a sphere with the same volume as the cell. {\em Bottom}: distribution of cell masses, plotted as a function of density. The colours are linearly related to the total number of cells. Most of the computational effort is spent in the high density regime where the Jeans-length criterion determines the cell masses.
    Here we reach sub-parsec resolution at densities comparable to the sink formation threshold. The Jeans-length criterion is responsible for the knee in the plots at a density of $\sim 5\times 10^{-23} \: {\rm g \, cm^{-3}}$.}
    \label{fig:resolution}
\end{figure}

\begin{table}
	\centering
	\caption{Mass resolution and softening length of the different particle types.}
	\label{tab:res}
	\begin{tabular}{lcr}
		\hline
		 & Mass resolution (M$_\odot$) & Softening (kpc)\\
		\hline
		Dark matter 	  & $6 \times 10^5$  & $0.2$  \\
		Stellar particles & $5 \times 10^4$  & $0.1$  \\
		Companion galaxy  & $4 \times 10^{10}$ & $3$  \\
		\hline
	\end{tabular}
\end{table}

The mass resolutions and softening lengths used for the different types of collisionless particles included in our simulation are listed in Table \ref{tab:res}.

Given the method described in the previous section, the gas cells start with an initial mass equal to the star particles from which they are drawn. The code is however able to quickly refine them during the first few million years of the simulation until the nominal resolution is reached. For our production simulation we set a target mass resolution for the gas cells of $300$~M$_\odot$. In the denser parts of the ISM, however, we reach considerably higher resolutions, down to a few solar masses (see Fig.\ \ref{fig:resolution}), since we require the local Jeans length to be resolved by at least four resolution elements in all gas with a density $\rho < \rho_c = 10^{-21} \: {\rm g \, cm^{-3}}$. This requirement generates a differential distribution of cell masses as a function of density and temperature. We reach high spatial resolutions in the star-forming part of the ISM despite a relatively low resolution in the more diffuse phase. For stability reasons, we also try to avoid having neighbouring cells with a large volume difference. If two neighbouring cells approach a volume ratio greater than 8, the larger cell is split. The resulting spatial resolution as a function of the local gas density and the corresponding mass distribution of the cells in the different density regimes is shown in Fig.\ \ref{fig:resolution}.

To help us to quantify the resolution dependence of our simulations, we have also carried out low resolution runs with a target mass resolution of $1000$~M$_\odot$. In these runs, we apply the Jeans-length refinement criterion only up to a limiting density of $\rho_{\rm lim} = 0.1 \times \rho_c = 10^{-22}$~g~cm$^{-3}$. This will result in a difference of spatial resolutions in the dense gas of about a factor of two with respect to the high resolution case. A comparison between the two simulations for a resolution study is therefore meaningful (see Appendix \ref{sec:resolutionDependece}).

Since the gas cells have different masses and sizes, we cannot use a unique gravitational softening length. Rather, we use {\sc arepo}'s adaptive softening option to scale the softening length according to the cell radius, i.e.\ $\epsilon_{\rm gas} = 2 r_{\rm cell}$, where $r_{\rm cell}$ is the radius of a sphere with the same volume as the cell.\footnote{As {\sc arepo} endeavours to prevent its grid cells from becoming highly distorted, most are quasi-spherical and so this radius is an accurate way of characterizing the size of the cells.}

If we compare the spatial resolution that we achieve in gas at typical GMC densities (a few times $10^{-22} \: {\rm g \: cm^{-3}}$ and above) with the requirements that were recently shown by \citet{Joshi+2019} to be necessary for producing converged molecular fractions in 3D simulations ($\Delta x \sim 0.2$~pc for H$_{2}$, $\Delta x \sim 0.04$~pc for CO), then we see that the chemical state of our simulations is not completely converged. Although we have more than enough resolution to successfully identify molecular-dominated clouds (the ``physical'' condition of \citealt{Joshi+2019}), we do not resolve the dense substructure within these clouds in enough detail to ensure that the molecular formation time is shorter than the cell crossing time in the densest cells (the ``dynamical'' condition of \citealt{Joshi+2019}). Therefore, although the details of the molecular gas distribution in our simulations should be qualitatively correct, some of the quantitative details may still be resolution-dependent
(see also Appendix~\ref{sec:resolutionDependece}).

\subsection{Simulation details}
\label{sec:Simulations}

\begin{figure}
	\includegraphics[width=\columnwidth]{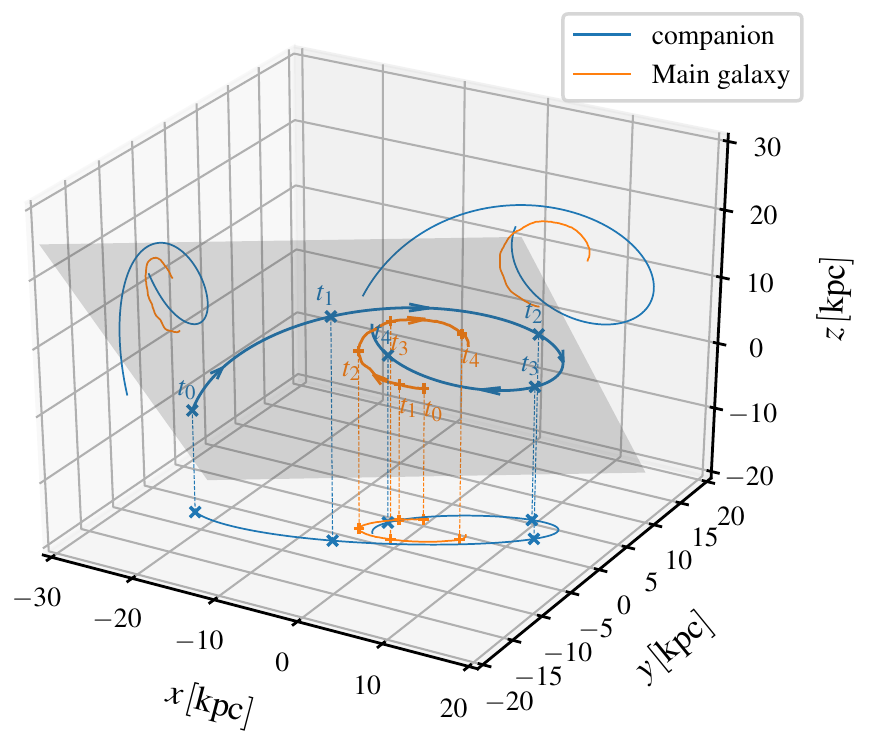}
    \caption{Trajectory of the centre of mass of the simulated galaxy ({\em orange line}) and its companion ({\em blue line}). The shaded area defines the plane of the orbit.}
    \label{fig:orbits}
\end{figure}

\begin{figure*}
	\includegraphics[width=\textwidth]{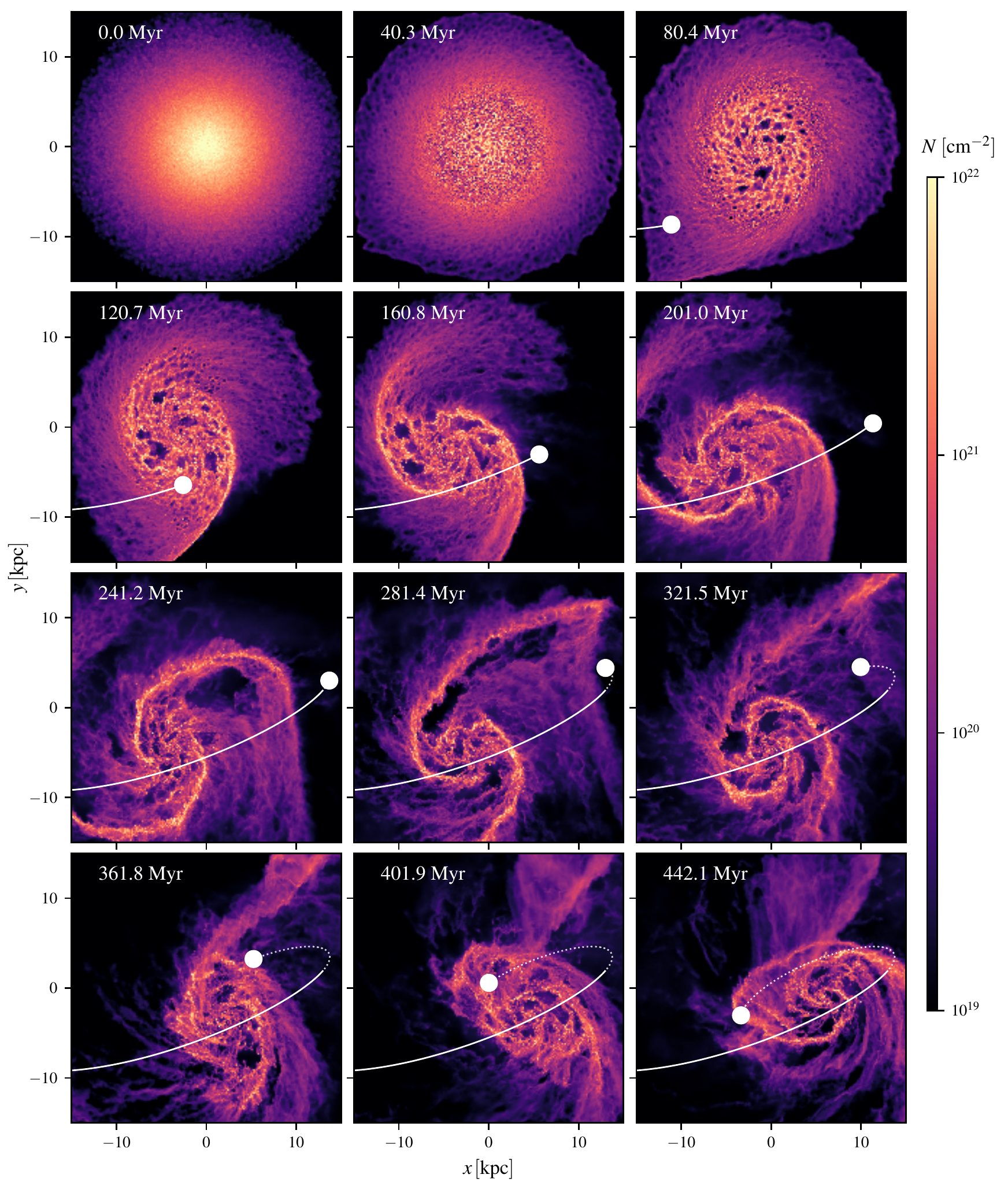}
        \caption{Total gas column density projections at different times for the interacting simulation. The location of the companion ({\em white filled circle}) and its trajectory ({\em white solid} and {\em dotted line}) are plotted in each panel. Notice how the interaction triggers a two-armed spiral structure. When the companion is behind the disc of the main galaxy the trajectory is plotted with a dotted line. The simulated galaxy and its companion in the last panel at $t = 428.9$~Myr are in a configuration similar to the observed M51 galaxy. The morphological changes induced by feedback can be clearly seen by comparing the last panel to the isothermal model shown in Fig.~\ref{fig:rhoproj_isothermal}.}
    \label{fig:rhoproj_v_time}
\end{figure*}

\begin{figure*}
	\includegraphics[width=\textwidth]{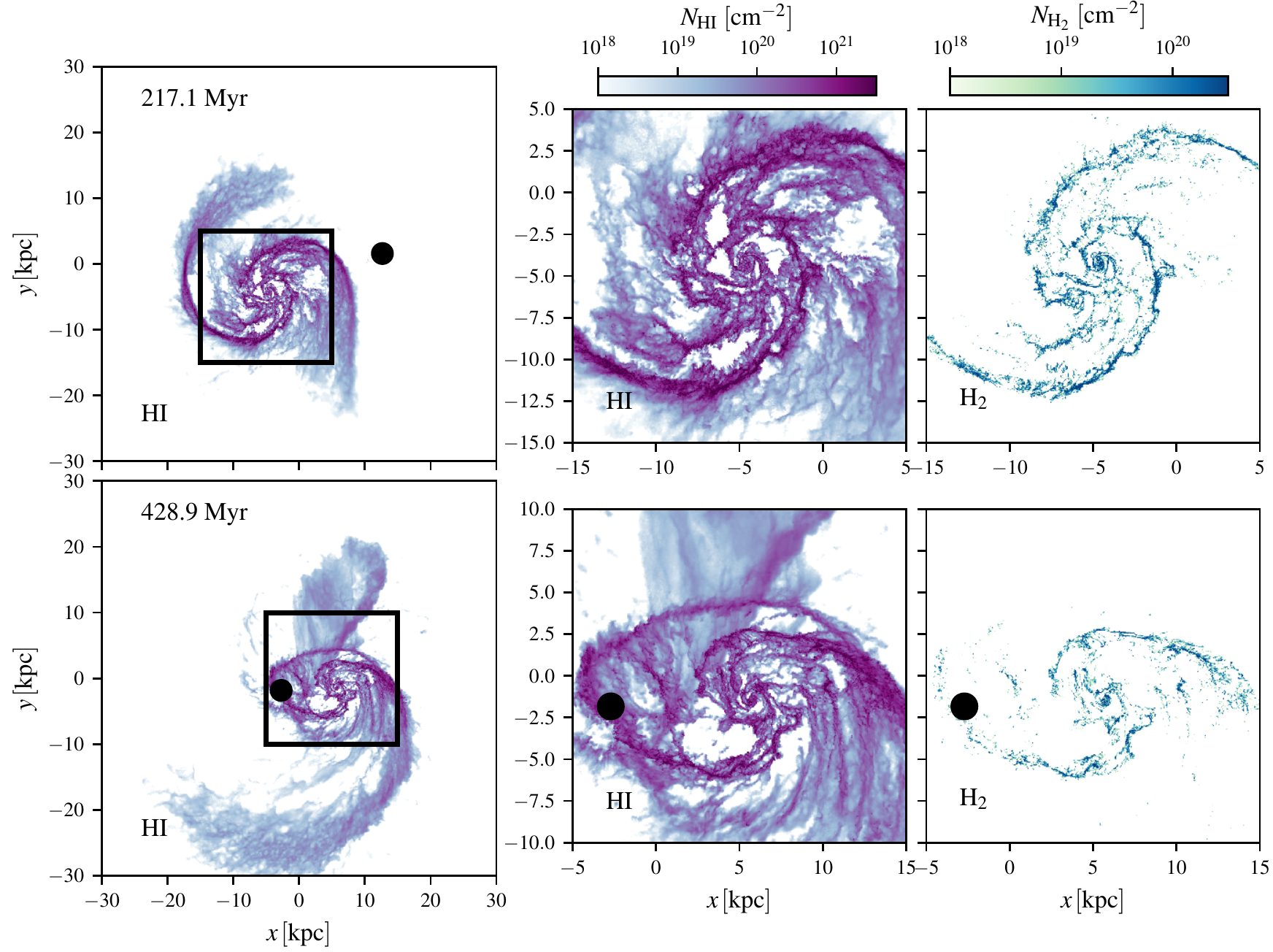}
    \caption{Column density maps of the interacting simulation at two different times. The left-hand panel shows maps of the H~{\sc i} column density at $t = 217.1$~Myr and 428.9~Myr. The remaining panels show a zoomed-in view of the H~{\sc i} ({\em central panels}) and H$_{2}$ ({\em right-hand panels}) column density in the central $20$~kpc of the galaxy at the same two output times. The grand design spiral arm pattern induced by the companion galaxy is clearly visible, especially in the upper panels. The pattern is much more pronounced in molecular gas. The bottom panels correspond to a configuration similar to the observed M51 galaxy. Also of note is the large tidal tail that has been ejected from the galaxy due to the close encounter with the companion, which is visible in the H~{\sc i} column density map but not in the H$_{2}$ column density map. The black dot indicates the position of the companion.}
    \label{fig:rhoproj_simulated_M51}
\end{figure*}

As stated above, our major focus in this study is to address the relative importance of the interaction in shaping the ISM. Therefore we set up two sets of simulations. In one case the galaxy is allowed to interact with a smaller companion galaxy as described in Section \ref{sec:IC}. The same galaxy is evolved in isolation in a comparison set of simulations. 

For both setups, we carry out calculations at two different resolutions, as described in Section \ref{sec:resolution}. Unless stated otherwise, all the analysis below refers to the high resolution simulations. The results of the other simulations are summarised in Appendix~\ref{sec:resolutionDependece}, where we discuss how our results depend on the chosen resolution.

For the high resolution simulations we simulate the first $\sim 40$~Myr at a low base resolution of $1000 \, \rm M_\odot$ per cell and without requiring the Jeans length to be resolved. We then switch to a base resolution of $300 \, \rm M_\odot$ per cell and finally at $\sim 80$~Myr we switch on the requirement for the Jeans length as described in Section \ref{sec:resolution} as well, reaching our nominal resolution as illustrated in Fig.\ \ref{fig:resolution}. With this gradual increment in resolution we can overcome the strong initial collapse due to the cooling of the atomic disc without investing too much computational power into this initial transition phase. 

\section{Results}
\label{sec:Results}
We start in this section by giving a general description of the outcome of the simulations, with a focus on the properties of the ISM and the SFR. We then look at how the interaction affects the total mass fractions in the different ISM phases and try to understand how the galaxy encounter influences the cold molecular gas reservoir which is available for SF.

\subsection{A qualitative description of the simulation}

\begin{figure*}
    \centering
    \includegraphics[width=0.95\textwidth]{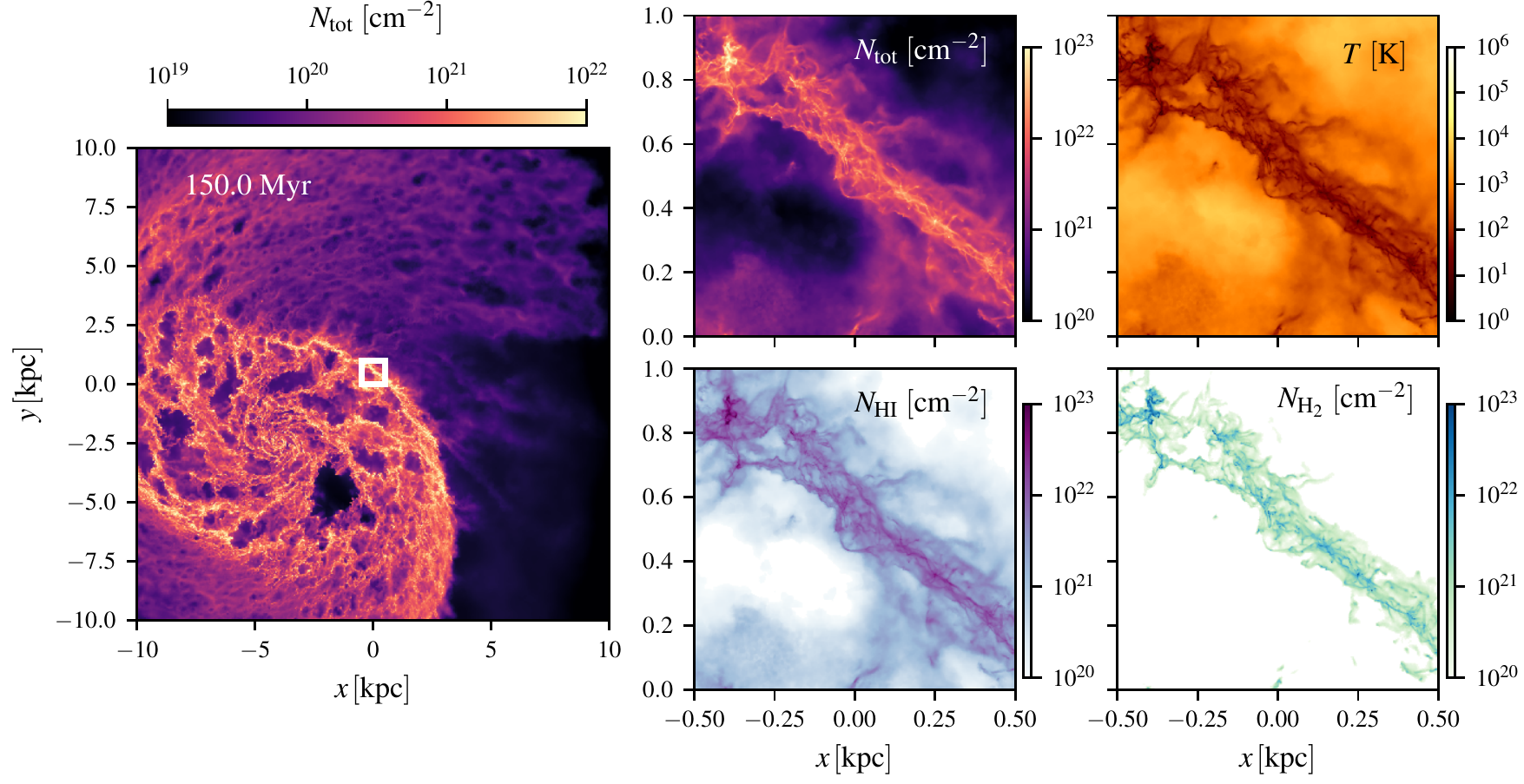}
    \caption{Total gas column density ({\em top left}), mass-averaged temperature along the line of sight ({\em top right}), H~{\sc i} column density ({\em bottom left}) and H$_2$ column density ({\em bottom right}) of a $(1\times1)$~kpc patch ({\em white square}) within the spiral arm of the interacting galaxy.}
    \label{fig:ProjZoom} 

    \includegraphics[width=0.95\textwidth]{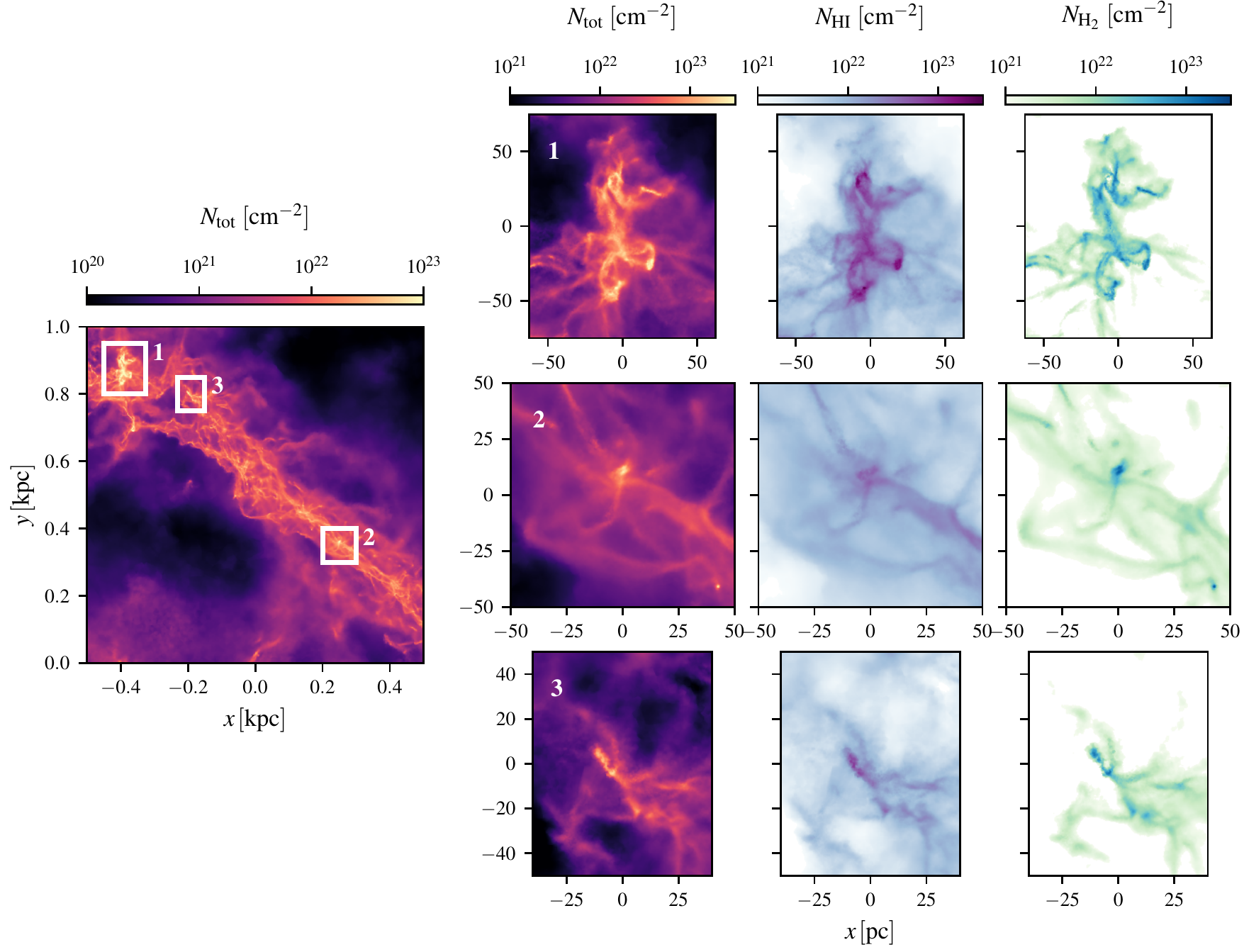}
    \caption{We further zoom in on a few selected GMCs of the spiral arm portion shown in Fig.~\ref{fig:ProjZoom}. We show the total ({\em left}), H~{\sc i} ({\em centre}) and H$_2$ column density maps for the selected regions.}
    \label{fig:ProjZoomOnClouds}
\end{figure*}

We follow the system for about $400$ Myr. This is the time when the relative position and morphology of the galaxy most closely resemble the observed M51 system. We show the trajectories of the main galaxy and its companion in Fig.\ \ref{fig:orbits}, while the evolution of the system in time can be followed looking at Fig.\ \ref{fig:rhoproj_v_time}. The companion galaxy moves on a highly eccentric orbit in front of the face-on disc relative to the observer and reaches its pericentric passage $\sim 110$~Myr after the start of the simulation. The companion then continues its orbit behind the disc of the galaxy at $\sim 270$~Myr. At the final snapshot the companion has a positive line-of-sight velocity with respect to the observer and is just emerging from behind the face-on disc of the main galaxy. At this point the distance between the two centres of mass is about $\sim 12$~kpc. Even though we stop the simulation at this time, the two galaxies will merge completely within the next orbit of the companion \citep[e.g.][]{Dobbs+2010}.

\begin{figure}
	\includegraphics[width=\columnwidth]{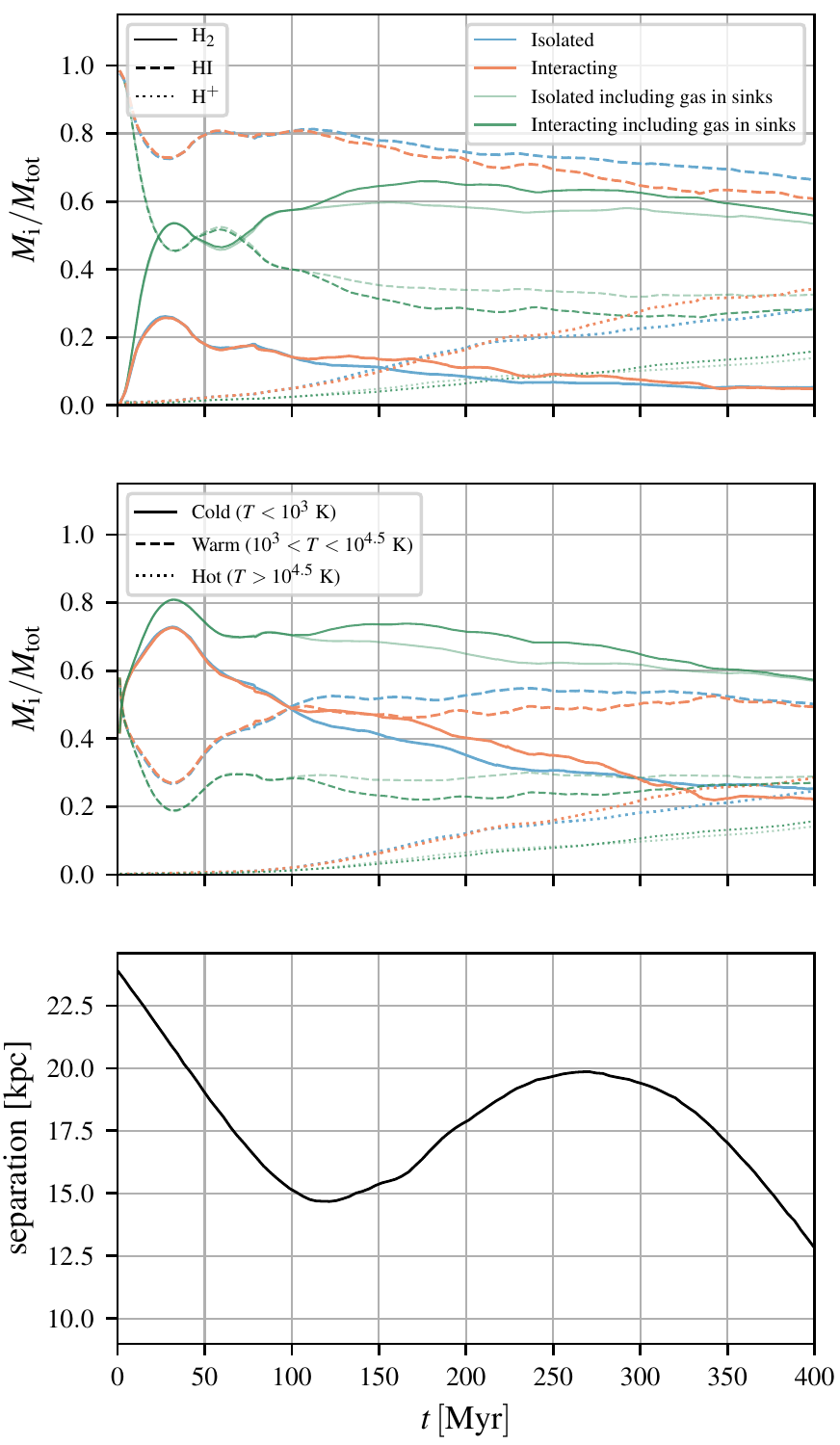}
    \caption{{\em Top}: Molecular {\em (solid line)}, atomic {\em (dashed line)} and ionised {\em (dotted line)} mass fraction as a function of time. Middle: Cold gas at $T<10^3$~K {\em (solid line)}, warm gas at $10^3<T<10^{4.5}$~K {\em (dashed line)} and hot gas at $T>10^{4.5}$~K mass fractions as a function of time. The interacting simulation is given in orange while the isolated one is depicted in blue. Here the gas trapped in sinks is not considered, but if we assume this gas to be cold and fully molecular the fractions change and are presented in green instead. {\em Bottom}: separation between the main galaxy's centre of mass and the companion as a function of time.}
    \label{fig:chem_phases_vs_time}
\end{figure}

\begin{figure}
	\includegraphics[width=\columnwidth]{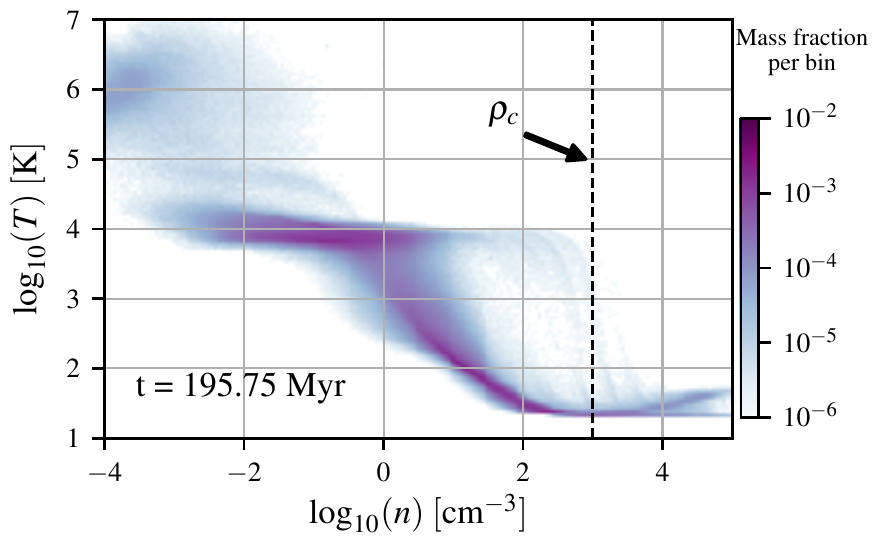}
    \caption{Density-temperature phase diagram of the gas phase in the interacting simulation. The colour indicates the total gas mass fraction in the given $(n, T)$ bin. The vertical dashed line indicates the density threshold for sink particle formation.}
    \label{fig:2DPDF_int}\end{figure}

\begin{figure}
	\includegraphics[width=\columnwidth]{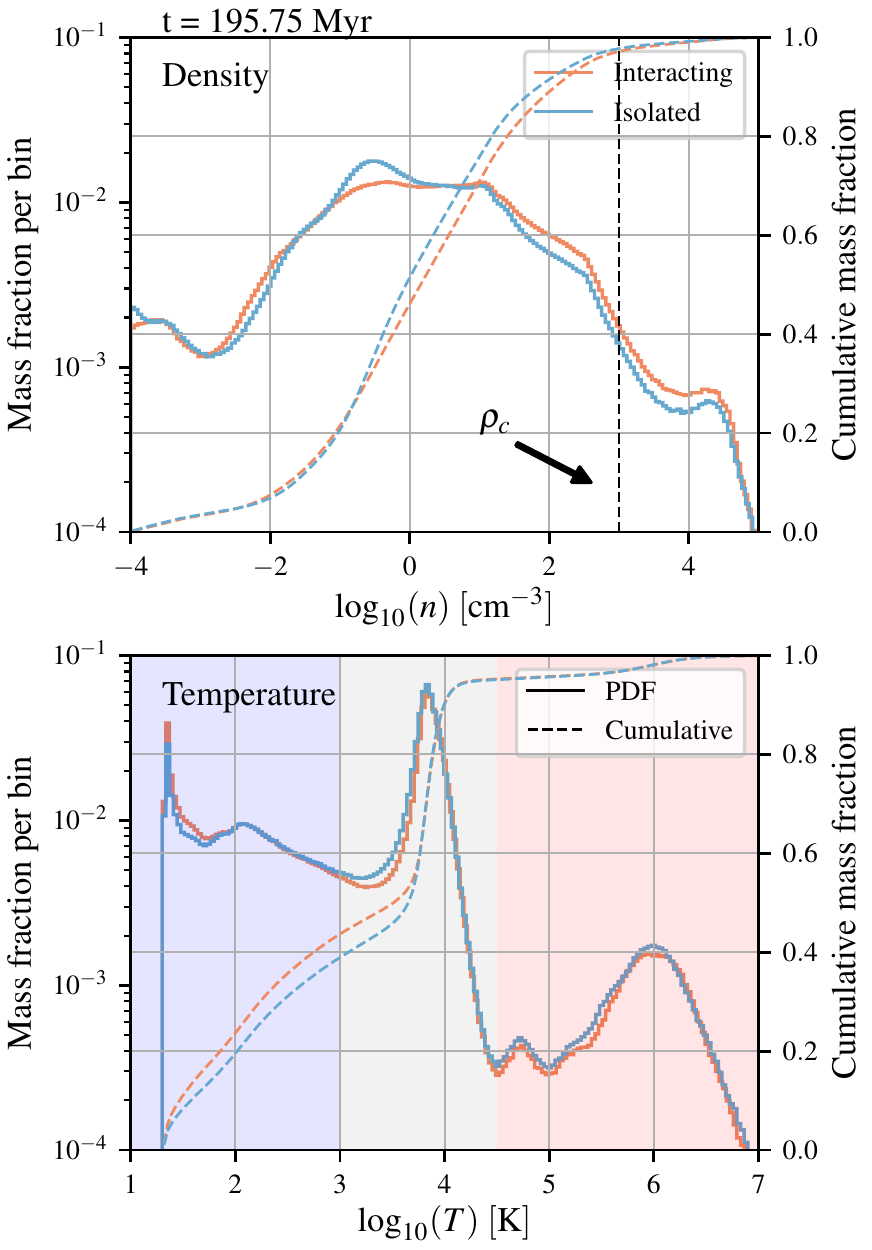}
    \caption{Mass-weighted density ({\em left panel}) and temperature PDF ({\em right panel}) of the gas phase in the simulations. The dashed line represents the cumulative density and temperature PDF respectively, i.e.\ the mass fraction with density/temperature below a given value. The blue line represents the isolated galaxy run while the orange shows the interacting simulation at the same time. The coloured regions in the temperature PDF indicate our definition of the cold/warm/hot phases (see text). Note that there is very little difference in the two cases highlighting how the interaction has little effect on the thermal phases of the ISM.}
    \label{fig:RhoTPDF}
\end{figure}

\begin{figure}
	\includegraphics[width=\columnwidth]{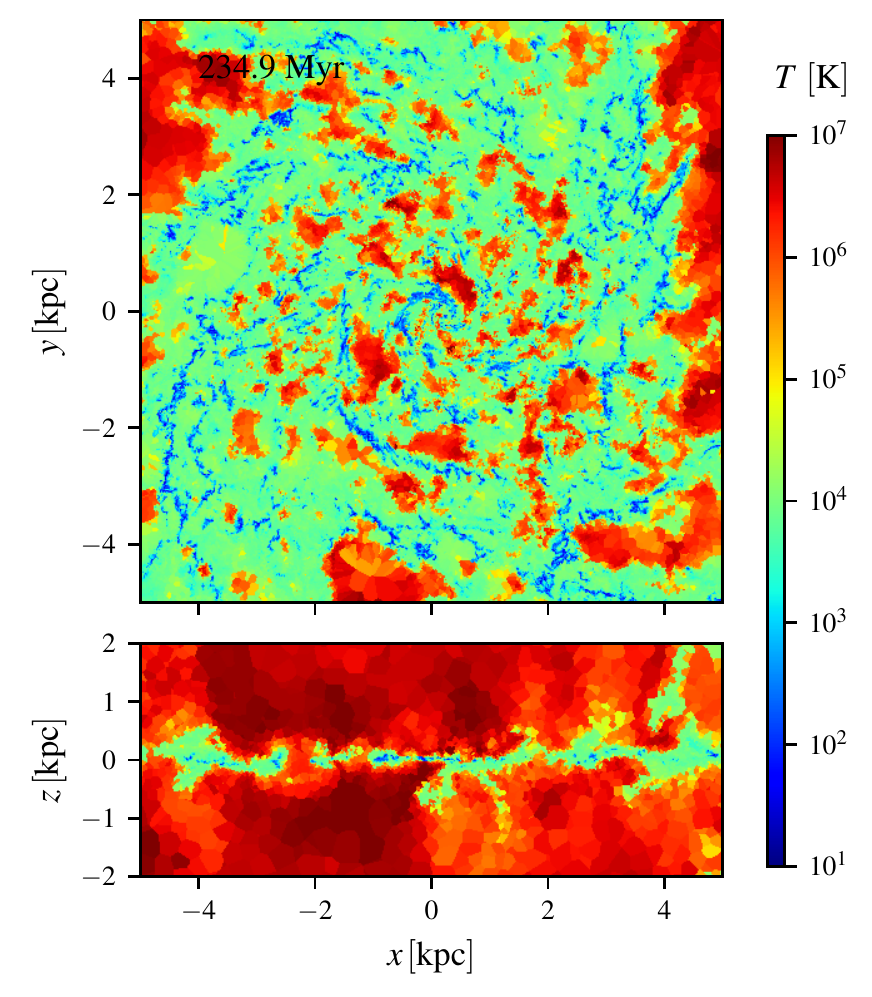}
    \caption{Temperature distribution in a slice in the $z=0$ plane ({\em top}) and $y=0$ plane ({\em bottom}) for the isolated simulation.}
    \label{fig:temperature_slices}
\end{figure}

The interaction is responsible for the development of a typical two-armed tidally induced spiral pattern in the disc \citep{Toomre&Toomre1972}. These arms are particularly pronounced in molecular gas and are the loci of intense SF (see Fig.\ \ref{fig:rhoproj_simulated_M51}). Due to the close passage, the outer parts of one of the arms are flung out, creating an extended tidal tail similar to what is seen in 21~cm observations of M51 \citep[see e.g.][]{Rots+1990}. This tidal tail is predominantly atomic as we can see in the bottom panel of Fig.\ \ref{fig:rhoproj_simulated_M51}. 

Due to gravitational instabilities the warm gas is pushed out of its thermal equilibrium and rapidly cools from its initial temperature of $T=10^4$~K. Part of it becomes molecular and builds up large and dense GMC associations. In the isolated galaxy simulation these are distributed in a flocculent style pattern, while in the interacting case they are mainly assembled inside the tidally induced spiral arm structure (Fig.\ \ref{fig:ProjZoom}). The clouds formed are filamentary, with complex substructure (Fig.~\ref{fig:ProjZoomOnClouds}), due to turbulence induced by a combination of SN feedback, galactic shear, and self gravity. See Smith et al. (submitted) for a discussion of the relative importance of the galactic potential with respect to the SN feedback in shaping the filamentary properties of clouds in a similar setup to the one used here. A detailed analysis of the GMCs in this M51-like galaxy simulation is deferred to a future study.

The ultraviolet component of the interstellar radiation field cannot penetrate these clouds and the pressure can thus drop quite substantially due to runaway cooling down to the temperature floor of $20$~K. This favours local collapse, which leads to intense SF. These newly-formed stars are responsible for clustered SN feedback that disrupts the parental clouds, creates large expanding superbubbles, and drives turbulence in the ISM. This favours a self-regulating matter cycle of the ISM \citep[see e.g.][]{MacLow&Klessen2004, Klessen&Glover2016} whose properties converge to a roughly steady state after $\sim 100$~Myr in the isolated case (see next paragraph). Only the slow depletion of gas affects this equilibrium. 

Since we lack early feedback such as winds or ionizing radiation, the earliest SNe exploding in each star-forming region are predominantly located in high density environments. In general, the momentum deposited by these SNe can create low density bubbles in which further SNe explode, with the combined effect of the clustered SNe eventually destroying the cloud. However, we encounter a few cases where this does not occur, so the SNe cannot completely disrupt the cloud. These pathological clouds continue to accrete mass onto sinks for a substantial amount of time, leading to extremely compact and massive clusters. We will come back to this point in Section \ref{sec:caveats}.

\begin{figure*}
	\includegraphics[width=\textwidth]{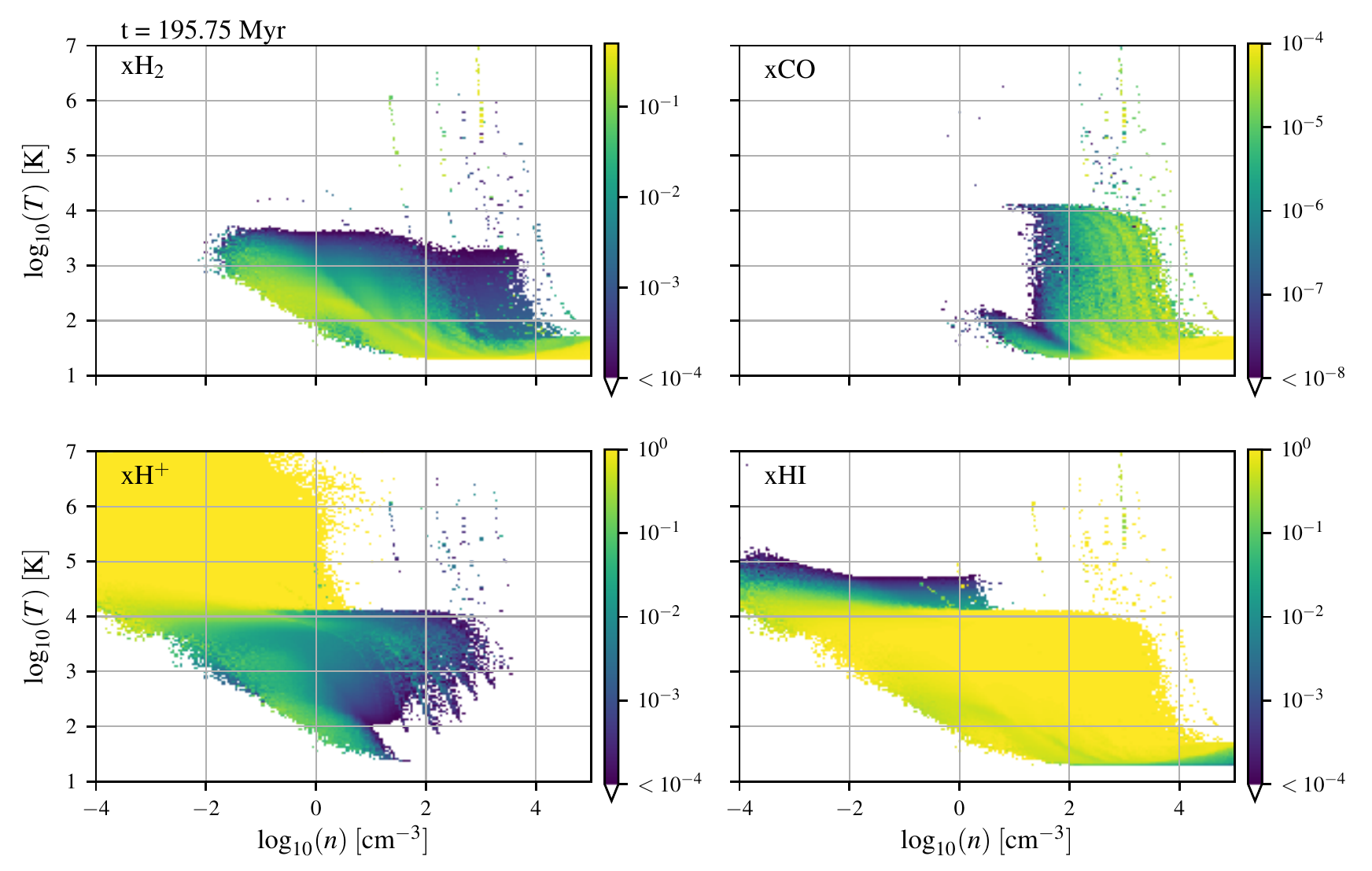}
    \caption{Mass-weighted average of the chemical abundance relative to the number of hydrogen nuclei in a given density--temperature bin for molecular hydrogen ({\em top left}), CO ({\em top right}), H$^{+}$ ({\em bottom left}) and atomic hydrogen ({\em bottom right}). Note that for molecular hydrogen, a fraction of $0.5$ corresponds to fully molecular gas.}
    \label{fig:ChemFractionInPhases}
\end{figure*}

\subsection{Thermal phases of the ISM}

Sink particles are formed in the densest collapsing parts of the cold ISM. As described in Section \ref{sec:SinkParticles}, this removes dense cold gas from the hydrodynamic simulation and locks it into collisionless particles. Only part of this gas is directly converted into stars, while the rest is temporarily trapped inside the sink and will be returned to the gas phase after the associated SNe occur. This prompts the question of how to account for this material when examining the distribution of gas across the different phases of the ISM, since we do not have any information about the density, temperature, or chemical composition of the trapped gas. One reasonable assumption would be that it is cold and fully molecular. However, this is likely an oversimplification; the real thermodynamical state of the gas could differ, especially if the stellar population within the sink is already in an advanced state of evolution and feedback has started to affect its contents. Nevertheless, this assumption does at least offer an upper limit on the cold molecular gas fraction in the simulations. Alternatively, by not including the gas in sinks at all, we instead recover a lower limit on the cold molecular gas fraction. This approach has the advantage that the state of the gas included in the analysis emerges self-consistently from the simulation and no additional assumptions have to be made. In our thermal analysis of the ISM phases, we have chosen the latter option and hence do not account for the trapped gas; the absolute value of the PDF and molecular fractions will be affected by this. However, this is the case for both isolated and interacting simulations, so a comparison between the two is consistent and the effect of the interaction can still be effectively studied. Finally, note that in some of the analysis later in the paper, we do attempt to account for the trapped material, making the assumption that it is cold and molecular. 

After an initial transition phase, shown in Fig.\ \ref{fig:chem_phases_vs_time}, the ISM in the simulated galaxy develops a three-phase thermal structure with a sharp lower limit of $20$~K due to the imposed temperature floor (see Fig.'s\ \ref{fig:2DPDF_int} and \ref{fig:RhoTPDF}). An upper limit of roughly 70\% of the gas mass is in the cold ($T<10^3$~K) phase, assuming the gas in the sink particles is primarily cold (Fig.\ \ref{fig:chem_phases_vs_time}), a fraction rather higher than the 50\% found in the Milky Way \citep{Ferriere2001}. This is perhaps appropriate for this actively star-forming galaxy.

Most of the remaining gas mass is in the warm stable phase around $T=10^4$~K with less than $10$~\% in the hot phase peaking at $10^6$~K, generated by SN feedback and strong shocks. Note however that our resolution prescription (Section \ref{sec:resolution}) is tuned to have the highest resolution in the dense part of the ISM at GMC scales, so it is likely that the fraction of gas in the hot diffuse phase where spatial resolution is small is not numerically converged. Numerical diffusion across interfaces in unresolved regions between hot and warm gas tends to favor cooling, thus under-predicting hot gas fractions. Moreover, a higher number of SN events with an unresolved Sedov-Taylor expansion implies a lower production rate of hot gas since momentum energy injection is unable to contribute to the hot phase. For an indication of the actual trend of the hot phase fraction as a function of resolution see Appendix \ref{sec:resolutionDependece}.

In Fig.\ \ref{fig:temperature_slices} we show the temperature of the gas in a slice through the mid-plane of the central $5$~kpc of the isolated galaxy. We can see that the cold phase is organised into GMC structures with a relatively low volume filling factor. These clouds are embedded in the warm phase at $T=10^4$~K while the clustered feedback coming from the sink particles drives superbubbles generating outflows producing the volume-filling hot phase, which permeates most of the volume above and below the disc.

Figure \ref{fig:ChemFractionInPhases} shows that only the densest and coldest parts of the ISM ($n > 10^2$~cm$^{-3}$ and $T\simeq 20$~K) are fully molecular and CO-bright. Because no CO is found below $n\lesssim 100$~cm$^{-3}$ the molecular gas in the transition zone between $\sim10$ and $10^4$~K is CO-dark.

In the top right panel of Fig.\ \ref{fig:ChemFractionInPhases} there are parts of the temperature-density phase space populated with high CO fractions but very little H$_2$. These are rapidly evolving SN shells where the simple model of CO chemistry used in our simulations does not capture the correct behaviour of the CO. The NL97 network used in our simulations assumes that photodissociation is the dominant destruction process for CO. In most of the ISM, this is a good assumption, but it breaks down in dense gas heated to high temperatures by strong shocks, where we would expect collisional dissociation of CO to dominate. Fortunately, this limitation of the NL97 network is highly unlikely to have a significant effect on the dynamical behaviour of the gas, since the cooling in these conditions is dominated by atomic line cooling and is hence insensitive to the CO content. In addition, the actual fraction of mass in this region of density-temperature space is small, as Figure~\ref{fig:2DPDF_int} makes clear.

As expected given the microscopic cross-section for collisional ionization of hydrogen, gas with a temperature $T \gg 10^4$~K is fully ionised. The atomic gas, on the other hand, lives in a wide region of the density-temperature phase space. A considerable fraction of the atomic gas is found at temperatures $T \sim 10^{4}$~K or $T \sim 100$~K, corresponding to the warm neutral medium (WNM) or cold neutral medium (CNM), respectively. However, there is also a substantial amount of atomic hydrogen in the transition region between these two thermally stable regimes (see also Figure~\ref{fig:2DPDF_int}), as is observed in the Milky Way \citep{Heiles2001}. It should also be noted that as our simulations do not include the local effects of ionizing radiation from young stars, we underestimate the ionization rate in warm ($T \sim 10^{4} \: {\rm K}$) low density gas, particularly far above or below the galactic midplane. Therefore, in our simulations this material remains largely atomic, while in a more realistic simulation it would be more highly ionised. As we are primarily interested in the behaviour of the dense gas and the SFR in our simulated galaxies, this should not have a significant impact on our results.

In Fig.\ \ref{fig:RhoTPDF} we also compare the density and temperature distributions of the ISM in the interacting galaxy to the isolated one. We notice that the interaction increases only marginally the amount of cold gas and the ISM phases are in general not affected by the merger. In Fig.\ \ref{fig:chem_phases_vs_time} where we show the time evolution of the gas mass in the different thermal/chemical phases for the two simulations, we see that after $\sim 100$~Myr, when the companion reaches its point of closest approach to the galaxy, the tidally-induced two-armed spiral pattern starts to develop and this correlates with an increase in cold molecular gas. However this difference is relatively small, amounting to only a few percent throughout the simulation time. Secular changes in the gas fractions as gas is consumed in SF are far larger.

\subsection{Star Formation}
\begin{figure*}
	\includegraphics[width=\textwidth]{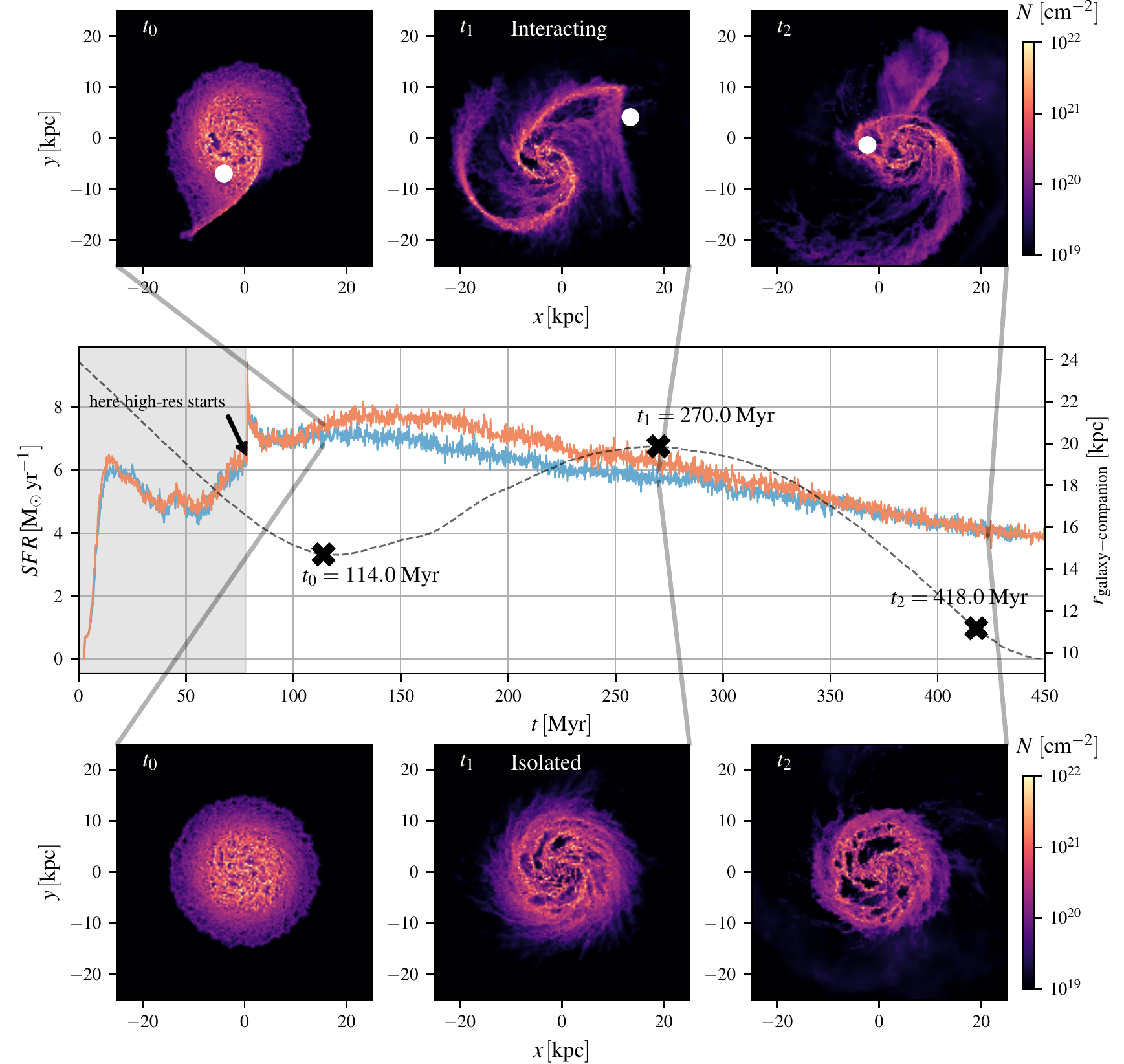}
    \caption{SFR as a function of time for the {\it Isolated} ({\em blue}) and {\it Interacting} ({\em orange}) runs, respectively. The black dotted line shows the separation of the centre of mass between the main galaxy and the companion for the interacting case. Three representative times are chosen ({\em black crosses}) for which we show the morphology of the interacting simulation ({\em top three panels}) and isolated simulation ({\em bottom three panels}). Despite the clear differences in the morphology of the galaxies in the two simulations, the SFR is surprisingly similar. This shows that the interaction has little effect on the SFR. The interaction merely dictates the morphology of the star-forming regions, but the intensity is controlled by the self-regulated feedback within the ISM.}
    \label{fig:SFR_proj}
\end{figure*}
Gravitational collapse occurs in GMCs where densities are highest and the ISM is cold enough to trigger runaway collapse leading to SF. We therefore expect a similar behaviour of the SFR to that of the cold molecular gas that we described in the previous section. 

\begin{figure}
	\includegraphics[width=\columnwidth]{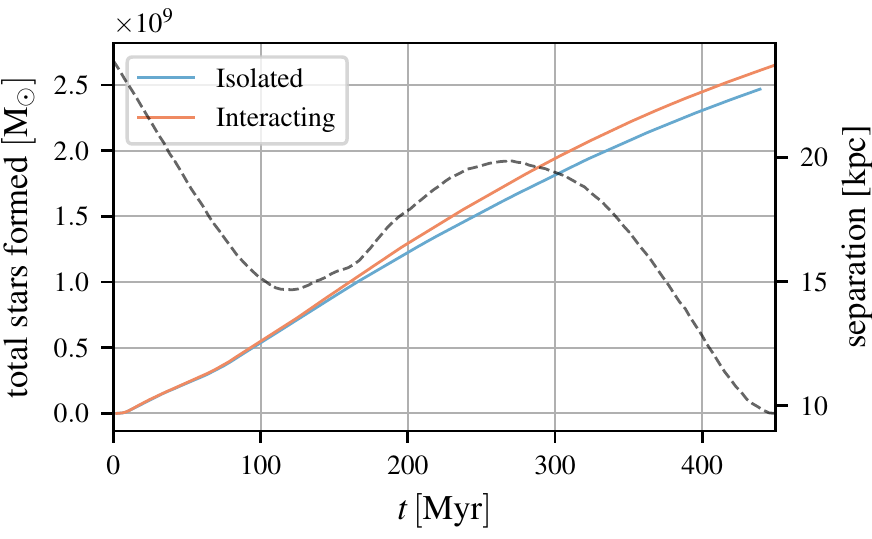}
    \caption{Cumulative SFR as a function of time, i.e.\ total amount of stars formed up to a given time $t$. The isolated galaxy is plotted in blue and the interacting one in orange. As in Fig.\ \ref{fig:SFR_proj} we also show the separation of the galaxy and its companion for the interacting simulation ({\em dotted line}).  }
    \label{fig:cumulative_SFR}
\end{figure}

We show the SFR as a function of time for the isolated and interacting runs in Fig.\ \ref{fig:SFR_proj}. Around $t = 80$~Myr we turn on the full refinement scheme and previously stable gas becomes unstable due to better resolved collapse. This explains the spike in SF noticeable just after the onset of our highest resolution scheme. In less than 20~Myr the ISM self-regulation brings this value back down to previous levels. SF increases after the point of closest separation, when the spiral arms start to develop ($t\simeq 110$~Myr). However, the difference between the isolated and interacting runs is small and at the end of the simulation the difference in the total mass of stars formed in the two simulations is less than $10$\% (see Fig.\ \ref{fig:cumulative_SFR}). 

At $t = 418$~Myr, the time at which our simulation is at an equivalent evolutionary phase to M51, we find an SFR of $4$~M$_\odot$~yr$^{-1}$ which is comparable to the observed value of $4.6$~M$_\odot$~yr$^{-1}$ \citep{Pineda+2018}. However if we consider that we started with a galaxy that had only half of the gas mass of M51a, we conclude that our depletion times are most likely too short by a factor of about two. This may reflect the lack of early feedback that can shut off SF earlier in the evolution of a newborn star cluster than SNe alone. 

\begin{figure}
	\includegraphics[width=\columnwidth]{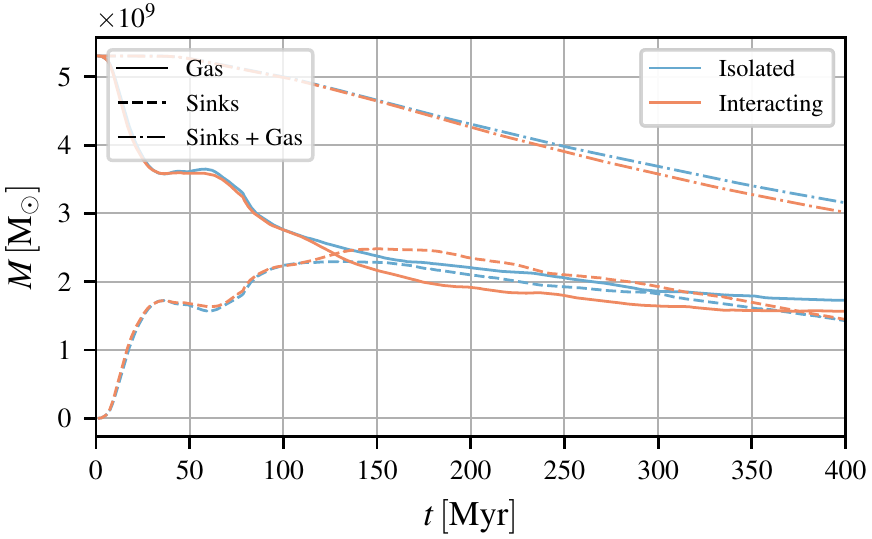}
    \caption{Gas mass ({\em solid line}) as a function of time for the interacting ({\em orange}) and isolated ({\em blue}) simulation. Part of the gas is locked into sink particles ({\em dashed line}) and will be returned to the gas phase over time, but part will be lost to stars contributing to a global steady depletion of gas. The dash-dotted line therefore shows the sum of the mass in sinks plus gas mass.}
    \label{fig:gas_depletion}
\end{figure}

We did not include any type of hot circumgalactic coronal gas from which the disc could replenish its gas reservoir, nor did we simulate other types of gas inflow. Therefore, even though we include a mass return from the sink particles, we are slowly depleting the gas available to SF (see Fig.\ \ref{fig:gas_depletion}). This is also reflected in the measured SFR in Fig.\ \ref{fig:SFR_proj}, where we see a slow but steady decline in SF at later stages of the simulation. 

To have an idea of where the SF is taking place, we look at how it correlates with the local gas column density. Observationally the gas surface density is connected to the SFR by a simple power law $\Sigma_{\rm SFR} \propto \Sigma_{\rm gas}^{\alpha}$ where $\alpha \simeq 1.4$ known as the Schmidt-Kennicutt relationship \citep{Schmidt1959, Kennicutt1989, Kennicutt1998}. An even narrower relationship with an exponent close to unity can be observed if only the molecular gas is considered \citep{Bigiel+2008}. Although this relation does not seem to be as universally applicable and in several instances can exhibit a large scatter \citep{Shetty+2014, Shetty+2014b}, it has been extensively used in the literature to connect large-scale galaxy properties directly to the local SF by abstracting the complexity of the SF process to a simple power law. 

\begin{figure*}
	\includegraphics[width=\textwidth]{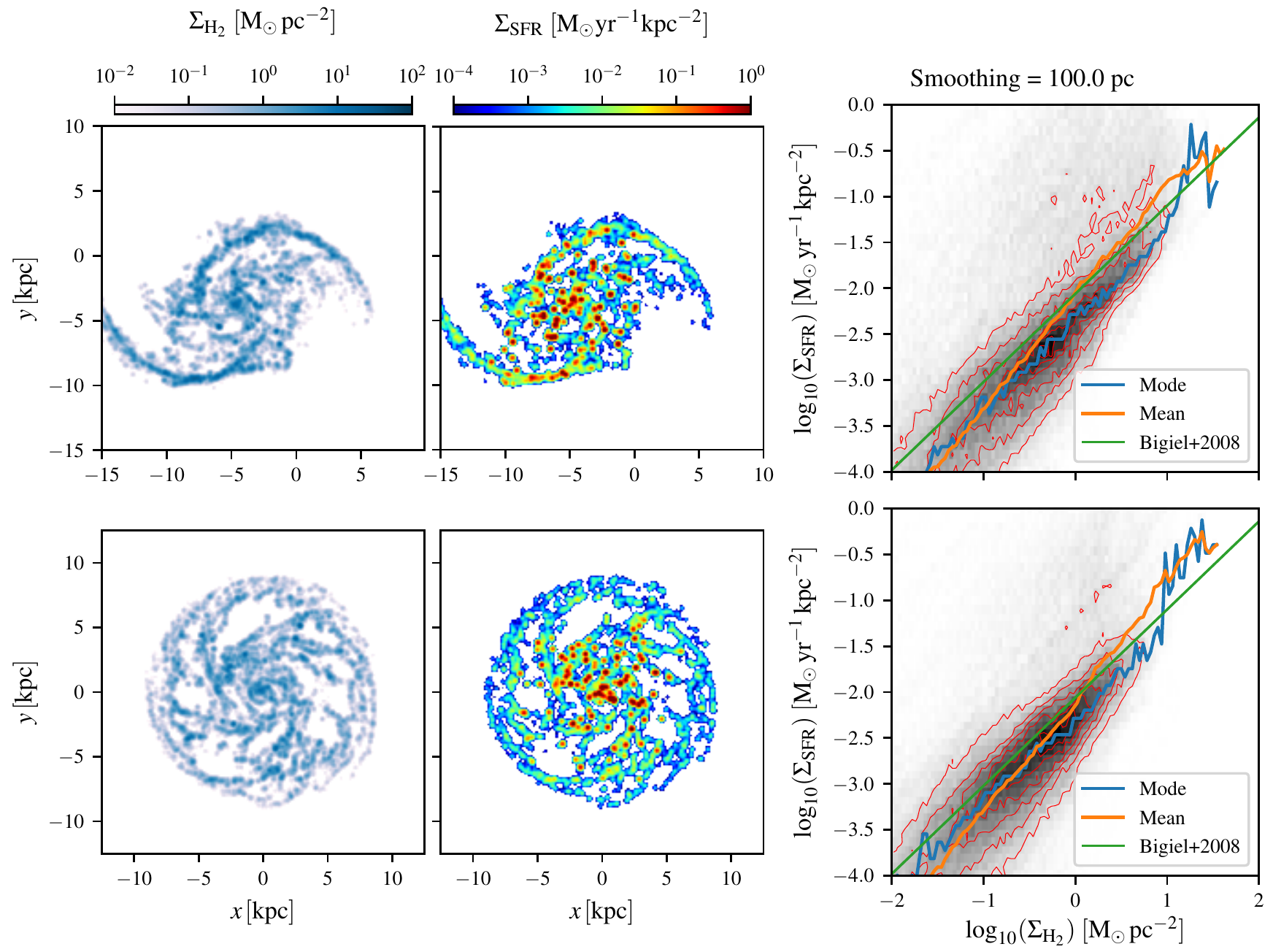}
        \caption{H$_2$ column density map convolved with a Gaussian function with standard deviation $\sigma = 100$~pc {\em (left)} at a time of $\sim 200$~Myr. SFR surface density map convolved with the same kernel {\em (centre)}. On the {\em right} we show the Schmidt-Kennicutt type relation based on the two maps. The {\em green} line is the observed relation taken from \citet{Bigiel+2008}, the {\em orange} line describes the mean while the {\em blue} line is the mode of the SFR distribution in each surface density bin. The {\em top} row indicates the interacting simulation while on the {\em bottom} we show the isolated one.}
    \label{fig:SK}
\end{figure*}

To see whether and how this relation develops in our simulations in Fig.\ \ref{fig:SK} we convolve the H$_2$ column density and the SFR surface density map with a Gaussian function of variable standard deviations $\sigma$ and then cross-correlate the two quantities. we find that the observed slope and scaling is well reproduced for $\sigma = 100$~pc. This smoothing is reasonable, as we do not expect molecular gas and SF to remain correlated down to arbitrarily small scales within galaxies \citep{Schruba+2010,Kruijssen+2018}, and a recent study has shown that the scale on which this decorrelation occurs is around 100--200~pc for a range of different galaxies \citep{Chevance+2019}. The mean SFR for every H$_2$ column density bin (orange line of Fig.\ \ref{fig:SK}), however, follows a slightly steeper power law with higher rates for higher surface density regimes with respect to the observed one. This connects to the low depletion times that we observe in our simulations (see next paragraph). 

We do not see a significant change in the slope, scaling and broadening of the relation if we consider the isolated galaxy simulation instead. This is an indication that the mechanism that controls SF is similar in the two instances and the relation emerges due to the local dynamics of the collapse, something that the galactic-scale interaction seems to have little impact on.

\begin{figure*}
	\includegraphics[width=\textwidth]{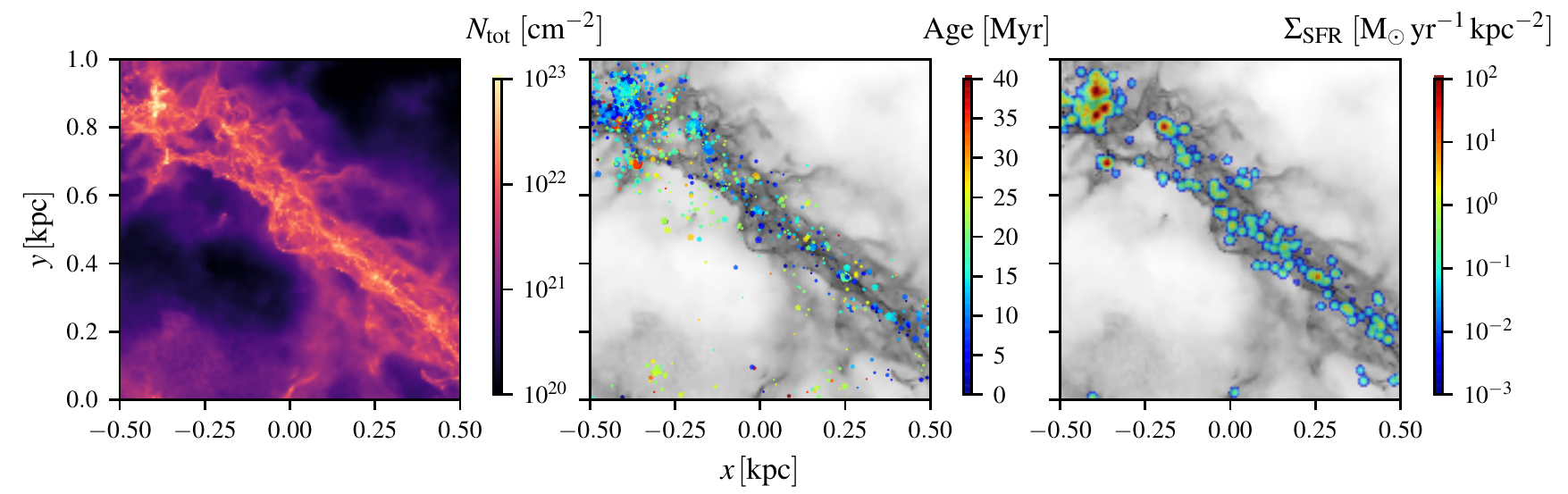}
    \caption{For the region shown in Fig.\ \ref{fig:ProjZoom} we show the location of the sink particles formed, coloured by their age ({\em central panel}) and the SFR surface density convolved with a Gaussian aperture of $\sigma = 5$~pc averaged over the past $4$~Myr.}
    \label{fig:ProjSFR}
  \end{figure*}

Morphologically, however, the two galaxies differ substantially and accordingly so does the distribution of the star-forming regions that are correlated with the molecular gas. The isolated galaxy remains flocculent throughout the simulation; frequently SF occurs at the edges of expanding superbubbles, which compress the gas and facilitate GMC formation. The interacting simulation instead develops strong spiral arms that correlate with SF (see Fig.\ \ref{fig:ProjSFR}). Since no new SF is generated during the encounter, it seems that the interaction is only grouping the GMCs and the associated SF into spiral arms as opposed to these structures being the trigger for new collapse. 

\section{Discussion}
\label{sec:Discussion}

\begin{figure}
	\includegraphics[width=\columnwidth]{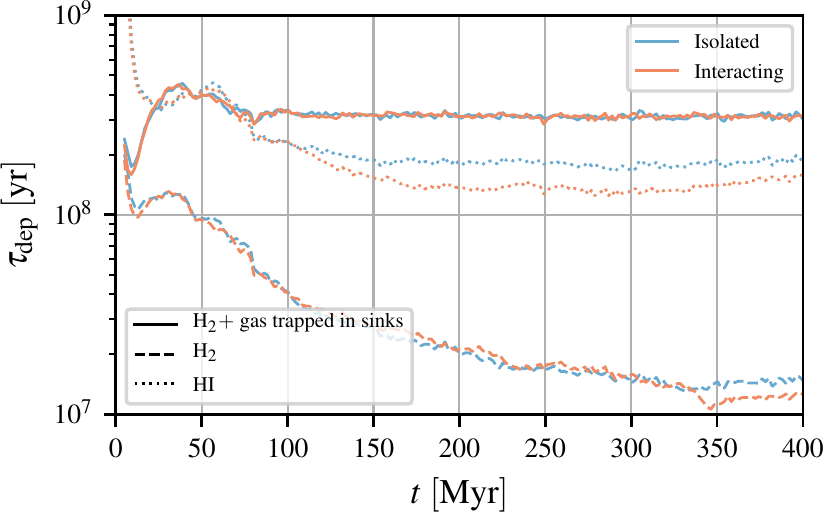}
        \caption{Depletion times ($\tau_{\rm dep} = M_{\rm i} / \rm{SFR}$) as a function of time for the isolated ({\em blue}) and interacting ({\em orange}) simulation. The H~{\sc i} ({\em dotted line}) and H$_2$ ({\em dashed line}) depletion times are shown. Since a lot of gas is trapped into sink particles, the H$_2$ depletion time is a lower limit. Considering all of the gas in the sink particles as molecular, we can sum it with the H$_2$ gas ({\em solid line}) to get an upper limit on the molecular gas depletion time.}
    \label{fig:depletionTime_vs_time}
\end{figure}

\begin{figure}
	\includegraphics[width=\columnwidth]{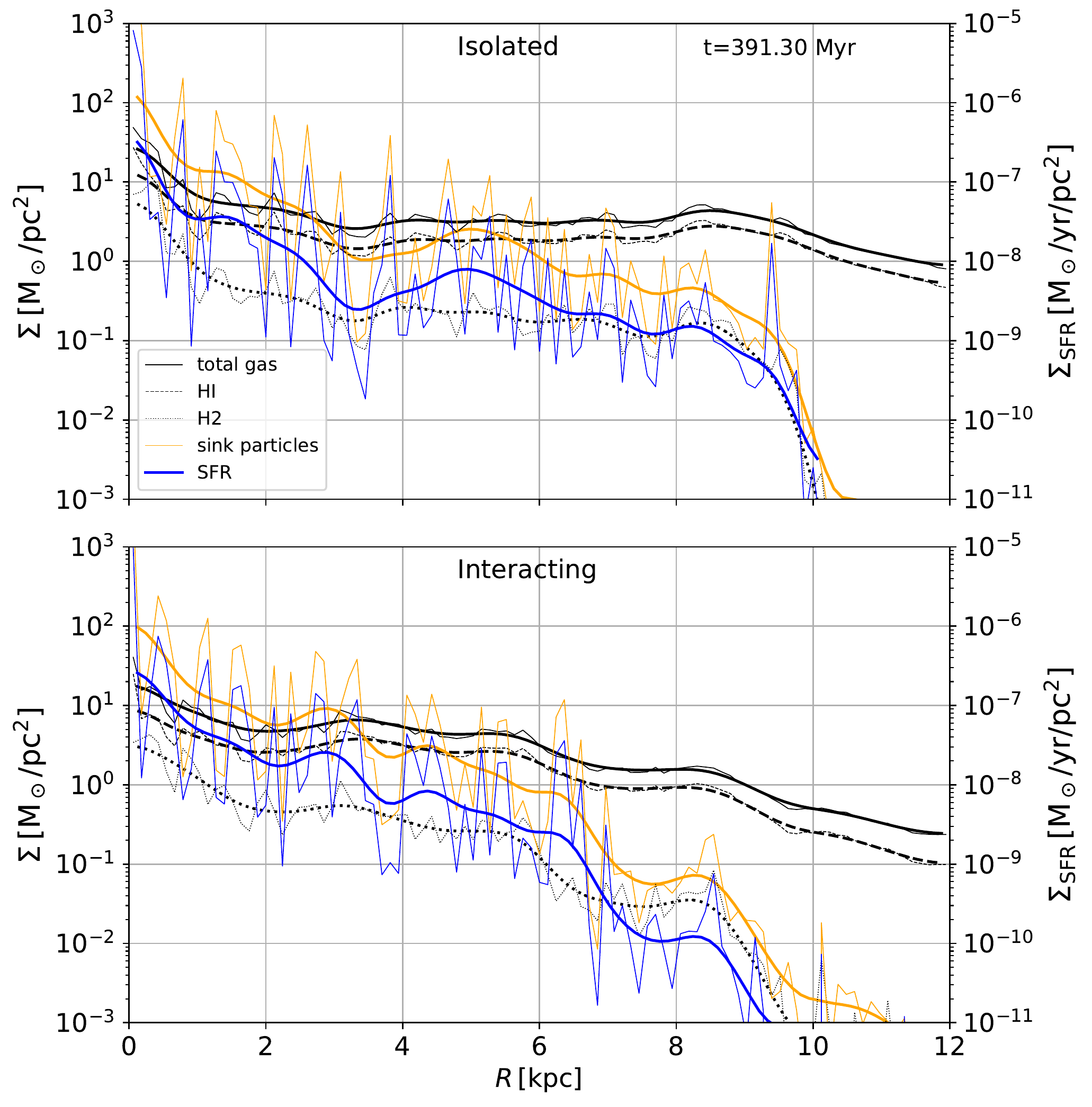}
    \caption{Surface density as a function of radius for the isolated {\em (top)} and the interacting simulation {\em (bottom)} at a simulation time of $\sim 400$~Myr. The total gas {\em (solid black line)}, H~{\sc i} {\em (dashed line)} and H$_2$ {\em (dotted line)} surface densities are shown. We also show the distribution of sink particles {\em (orange line)} and the SFR surface density {\em (blue line)}. Due to the high radial variability for some of the quantities, we also plotted {\em (thicker lines)} the same quantities convolved with a Gaussian filter to better show the radial profile.}
    \label{fig:surface_density}
\end{figure}

\begin{figure}
	\includegraphics[width=\columnwidth]{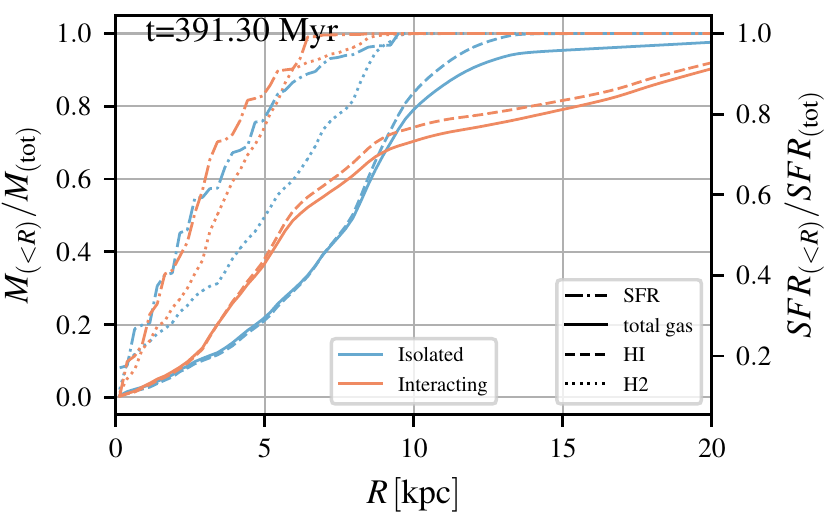}
        \caption{Cumulative total/H~{\sc i}/H$_2$ gas mass distribution {\em (solid, dashed and dotted lines)} as a function of galactocentric radius normalised to the total mass in each component for the isolated {\em (blue)} and interacting galaxy {\em (orange)} at a simulation time around $400$~Myr. We also show the normalised cumulative SFR as a function of galactocentric radius for the two simulations {\em (dot-dashed line)}. }
    \label{fig:cumulative_mass}
\end{figure}

\begin{figure}
	\includegraphics[width=\columnwidth]{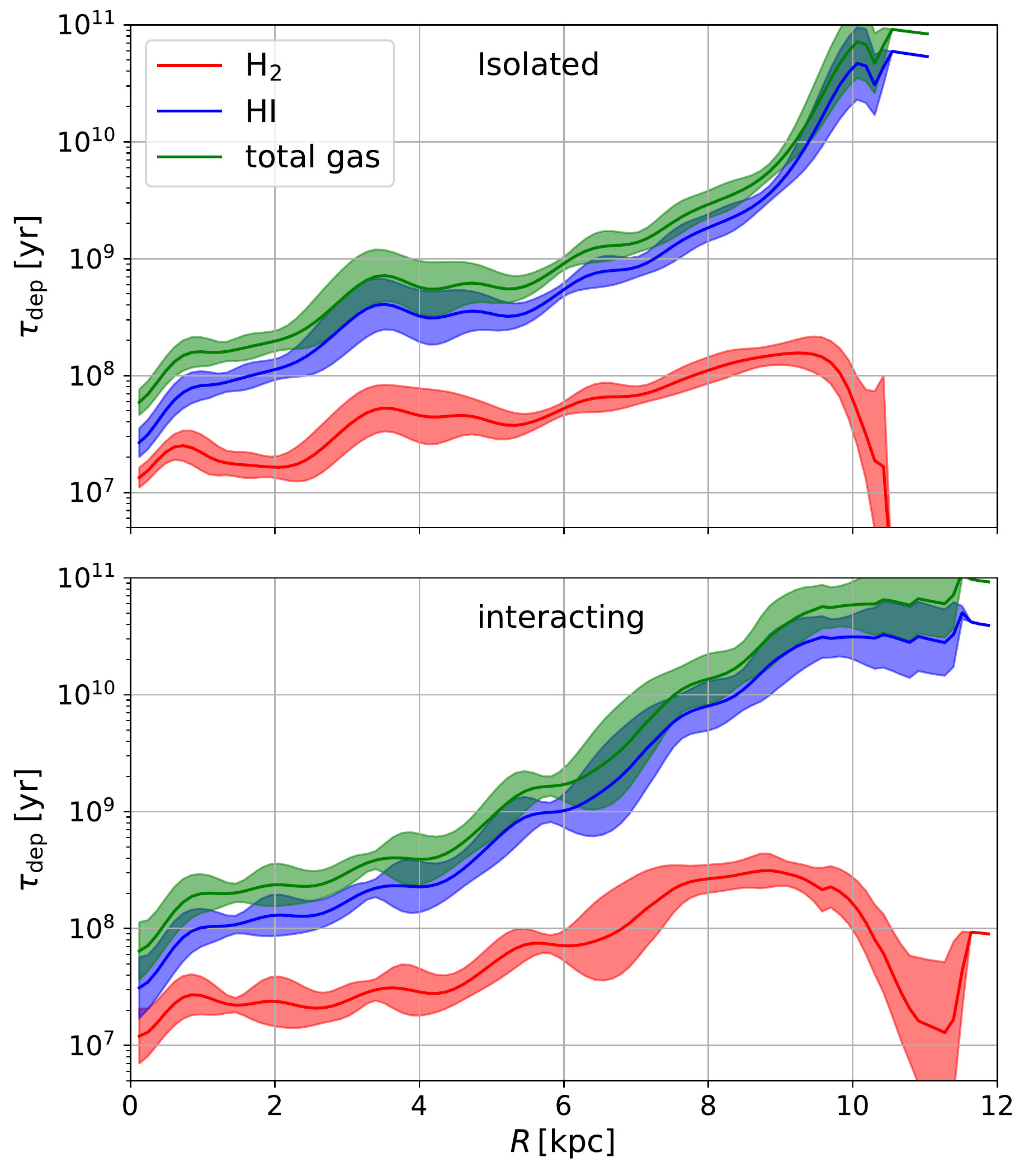}
        \caption{Depletion times of the total {\em (green)}, molecular {\em (red)} and atomic {\em (blue)} gas as a function of radius for the isolated {\em (top)} and interacting {\em (bottom)} galaxy. The depletion times are calculated for a simulation time around $400$~Myr and averaged over a time period of $\sim 20$~Myr. The shaded area is a one sigma deviation from that average.}
    \label{fig:depletion_time_vs_R}
  \end{figure}

Given the findings of the previous two sections, we see that the interaction, for the type of encounter considered here, is not able to significantly change the structure of the ISM in terms of thermal phases and chemical state (see Figs.\  \ref{fig:chem_phases_vs_time}, \ref{fig:RhoTPDF}). This is reflected in the almost identical SF history experienced by the two galaxies (Figs.\ \ref{fig:SFR_proj}, \ref{fig:cumulative_SFR}). 

Major galaxy interactions are generally associated with enhanced SFRs \citep{Larson&Tinsley1978, Lonsdale+1984, Barton+2000, Ellison+2008, Renaud+2019}. Outliers are however possible and the details are strongly dependent on the orbital parameters of the encounter and the stability of the isolated disc \citep{DiMatteo+2007}. This is actually in line with the inferred SF history of the M51 galaxy which lacks the fingerprint of enhanced SF activity but had a roughly constant rate of $5 \, \rm M_\odot yr^{-1}$ during the past few gigayears which even declined somewhat in the past $100$~Myr \citep{Eufrasio+2017}. This is roughly what we see also for the simulated system (Fig.\ \ref{fig:SFR_proj}.

Most simulated mergers, however, show the peak of the SFR during the coalescence phase \citep{Cox+2006, DiMatteo+2007, Renaud+2014}, which could be associated with a different behaviour of the ISM due to more extreme galaxy conditions. This phase is not followed by our model, so it is possible that the bulk of the SF is yet to come. 

The strength of the SF burst decreases for smaller mass ratios between the two galaxies and can be negligible when the tidal disturbance is small, as in the case of minor mergers \citep{Cox+2008}. Even though our nominal mass ratio turns out rather low, based on dynamical considerations we have argued that our simulated companion galaxy de facto represents just its core (see Section \ref{sec:IC}) and that the actual mass ratio is closer to the observed one of M51. If this statement is erroneous, on the other hand, the lack of enhanced SF is not an exception, but rather the normal behaviour for mergers in this mass regime in agreement with \citet{Cox+2008}. A better treatment of the mass distribution of the companion galaxy, instead of treating it as a simple point mass, will shed light on this. 

\citet{DiMatteo+2007} suggested that in some cases the close encounter can eject considerable amounts of gas into the diffuse tidal tail, which then cannot fully re-accrete at later stages of the merger, thereby explaining the lack of enhanced SF. Our simulation also develops such a diffuse atomic tidal tail (bottom panel of Fig.\ \ref{fig:rhoproj_simulated_M51}). It could therefore be that this is removing significant amounts of gas from the pool available to SF while still increasing the SFR in more central regions. 

From Fig.\ \ref{fig:cumulative_mass}, where we plot cumulative masses as a function of radius for the two simulations, we can see that this is however not the case. Almost 20\% of the total mass of the interacting galaxy is at $R\gtrsim 10$~kpc and thus in the tidal tail, however in the isolated case a similar mass fraction is in the part of the disc that is mainly atomic and not star forming. Therefore it is unlikely that the interaction is effective at removing gas which would otherwise have been available for SF. 

We conclude that the two galaxies have essentially the same amount of gas available for SF throughout the simulation. This is also seen in Fig.~\ref{fig:depletionTime_vs_time}, where we show the depletion times of the different ISM components versus time and notice that between the interacting and the isolated simulation there is only a marginal difference during their evolution. Globally, the interaction is not able to change the relation between the available gas and the SFR, which explains also why the inferred Schmidt-Kennicutt relations for the two cases are the same.

As suggested in Fig.\ \ref{fig:cumulative_mass}, we do not see a significant mass flow to the very centre, as the mass profiles up to $\sim 2.5$~kpc are similar in the two simulations. Other studies found that galaxy interactions drive large gas flows towards the central regions and therefore drive a nuclear star burst \citep{Torrey+2012, Moreno+2015}. This is not reproduced in our high resolution simulations for this specific merger system, and so the SF rate remains low.

The interaction is, however, able to redistribute the gas mass within the disc for more intermediate radii as we can notice from Fig.\ \ref{fig:surface_density}, where we plot the radial surface density profiles. As a matter of fact, in the interacting case roughly $60$\% of the mass is within the central $\sim 6$~kpc while the same radius contains only $30$\% in the isolated disc (Fig.\ \ref{fig:cumulative_mass}). Such a strong difference between the two is absent if we look at the cumulative SFR profile (dot-dashed line in Fig.\ \ref{fig:cumulative_mass}). This indicates that the interaction studied here is able to produce changes to the SF efficiency locally. This can also be appreciated in Fig.\ \ref{fig:depletion_time_vs_R} where we plot depletion times as a function of radius. We see here that variations of the order of a few are possible. These differences are significant as they are greater than just the temporal fluctuations of the local depletion times.

The question arises then as to what is controlling the ISM phases and the SFR if the interaction is ineffective in doing so. The isolated disc collapses and generates GMCs leading to SF in the central $8\text{--}10$~kpc which is the region that is initially marginally Toomre unstable (Fig.\ \ref{fig:vc}). Once the collapse started, a self-regulating equilibrium is generated where the energy input from the stellar feedback acts to counterbalance the forces responsible for cooling the gas to GMC levels to be available for SF again. The disc is essentially maximally star forming in the sense that in the Toomre unstable regions there is no gas reservoir that is not available for SF. Stars form at a rate set by the requirement that feedback balance the vertical pressure in the disc \citep{Ostriker+2010, Ostriker&Shetty2011}. The interaction is not able to increase the SFR since no new gas is added and all of the available gas is already available for SF. If the encounter is then not able to considerably change the conditions that control the turnover time of molecular gas, such as midplane pressure, SFR is unaltered. 

These conclusions are case specific for the type of interaction studied here. If for instance the companion galaxy had a non-negligible ISM fraction, direct collision of the two gas discs could have led to local collapse in the Toomre stable part of the isolated disc, resulting thus in an enhanced SFR.

Moreover, if the mass ratio between the two galaxies were greater, we might have seen a more pronounced mass flow towards the centre, more compression in the tidal tail, and an increase of midplane pressure, all factors that could lead to higher SFRs, either by changing the available amount of gas or by increasing the SF efficiency. Exploring these issues will require a more extensive parameter study of galaxy interactions, which is out of the scope of the study presented here.

\section{Caveats}
\label{sec:caveats}
The ISM is dynamically complex, and so our simulations are inevitably a simplification compared to the true behaviour of the ISM in a galaxy. In particular, there are several important physical processes that are not included in our current model. These include magnetic fields and early stellar feedback such as ionizing radiation and winds.

Early feedback is responsible for clearing out the surrounding gas so that when SNe explode, every SN event can deposit a much higher energy into the ISM than in the case of SNe being directly injected into the high-density molecular phase. In our simulations it is often the initial SN that takes over the role of early feedback of clearing out the surroundings and preparing the region for the later SNe to disrupt the cloud. Although the specific details of this disruption are likely sensitive to the specific feedback implementation, the sole fact of having a mechanism to self-consistently disrupt clouds from within ensures that the ISM is self-regulated by the internal feedback and so a healthy matter-cycle is achieved. These simulations are therefore well suited to address questions regarding the global life cycle of the ISM or the formation and early stages of GMCs. 

In a few cases, however, we find that the initial SNe explode in such a dense environment that they cannot efficiently pre-process the surrounding ISM for later SNe to be effective. Instead, in these cases the injected energy is quickly radiated away and the bubble re-collapses before further SNe can pressurise its interior and drive further expansion. Consequently, SF cannot be halted by feedback but instead continues for an unphysically long period of time, generating extremely massive star clusters and long-lived GMCs. While some models predict re-collapse of massive clouds and subsequent SF cycles \citep[see e.g.][]{Rahner+2017,Rahner+2018,Rahner+2019}, in our simulations this is largely a numerical artefact. 

If for some dynamical reason these massive clusters decouple from the parental cloud\footnote{The gas is collisional while the stars are collisionless, so it is not unusual for the two components to decouple, for instance during cloud collisions.}, the SNe associated with the cluster can deposit their energy much more efficiently into the ISM. Since the cluster is unphysically massive, it also produces a large number of SNe. The resulting superbubble can therefore be extremely large and have a significant impact on the morphology of the entire galaxy. This is probably a major reason for the spiral arms being much less defined in the interacting simulations compared to the isothermal runs (compare Fig.\ \ref{fig:rhoproj_isothermal} and Fig.\ \ref{fig:rhoproj_simulated_M51} for instance).

Early feedback is also responsible for shutting down SF much earlier in the life of a young GMC than in the case of SNe alone \citep[see e.g.][]{Gatto+2017,Kannan+2018,Fujimoto+2019}. The absence of early feedback may therefore lead us to overestimating the SFR. To some extent this has been corrected for by our assumption of the local SF efficiency within the individual sink particles (Section \ref{sec:SinkParticles}). Moreover, even if our SFRs are overestimated in some cases, the effect should be comparably strong in both simulations, meaning that the trend that we see in Fig.\ \ref{fig:SFR_proj} should be similar and our conclusions should not change. 

Our neglect of magnetic fields means that we are missing a source of stabilizing pressure against collapse and cloud formation. The compression of gas into spiral arms due to the galactic interaction could potentially be a trigger to overcome this additional pressure force and initiate cloud formation. In this case the interaction with the companion could have a more dramatic effect on the cold molecular phase than in the simulations presented in this paper. This should be investigated in dedicated studies. 

\section{Conclusions}
\label{sec:Conclusions}

We have performed high resolution {\sc arepo} simulations of a massive spiral galaxy interacting with a smaller companion. The properties of the galaxies and the orbital parameters of the encounter were chosen to roughly reproduce an M51-like system. For comparison purposes, we also modelled the evolution of the spiral galaxy in the absence of the interaction. Our simulations reach sub-parsec spatial resolution in dense molecular gas throughout the galaxy. We include the major physical ingredients thought to play key roles in the formation and destruction of GMCs to get a healthy life-cycle of the molecular gas in the galaxy. These include a time-dependent, non-equilibrium chemical network able to follow hydrogen and CO chemistry, local shielding from the molecule-dissociating part of the interstellar radiation field, sink particle formation to follow local centres of collapse and model stellar birth, and coupled SN feedback. 

The isolated galaxy stays mostly flocculent throughout the simulation while in the case of the interaction a strong two-armed spiral pattern develops, along with an extended atomic tidal tail similar to the one observed for M51a. The final morphology and configuration closely resembles the M51 system, although our feedback prescription created strong superbubbles disrupting the otherwise clean spiral pattern much more than in the real case. 

The ISM in the simulations settles into a typical three-phase medium with the cold molecular gas organised into dense GMCs associated with intense SF. Atomic gas makes up the cold neutral medium as well as the warm $T\sim10^4$~K phase into which the molecular cloud complexes are embedded. Supernova explosions coupled to recent SF activity are responsible for creating large superbubbles disrupting surrounding clouds and generating the hot ionised volume filling phase. The ISM properties roughly converge to an equilibrium state after an initial transition phase and only vary slowly after this due to gas depletion.

A lower limit of $\sim 10$\% and an upper limit of $\sim 60$\% of the gas mass is molecular, depending on what one assumes regarding the chemical state of gas trapped inside sink particles. We find an SFR of $4.0$~M$_\odot$ yr$^{-1}$ at a time corresponding to the current evolutionary phase of M51, in good agreement with the measured value of $4.6$~M$_\odot$ yr$^{-1}$. Due to lack of early feedback, however, our depletion times are too low by a factor of at least two considering that we started with a less massive ISM disc.

With this study we tried to further understand how the interaction of two galaxies can affect the ISM and the resulting SFR. Galaxy interactions are frequently invoked to induce star-bursts and to produce a general increase in SFRs. In the case analysed here, however, we find that other factors such as the initial disc stability and local feedback are more important than the interaction itself for controlling the ISM properties. While morphologically very different from each other, we find that the ISM phases of the two simulations are only marginally affected by the interaction, resulting in an almost identical SF history for the two cases. The galaxy interaction is {\em not} the trigger of strong star-bursts in the disc for our simulations. The M51 system is therefore a prototypical example of a merger event where SF is not controlled by the interaction but rather by pre-existing galaxy conditions and the self-regulating nature of the ISM. This is also supported by the observations that suggest a roughly constant SF rate during the past several $10^8$ years.

In the two scenarios simulated, the total gas accessible to SF is roughly the same. The interaction can not remove gas from the pool available to SF by shooting it into the diffuse tidal tail, as that mainly comes from an already stable part of the isolated disc. Nor is it compressing previously stable gas in the outskirts of the disc enough to trigger additional SF there. The global depletion times are therefore very similar in the two simulations. Locally, however, the interaction modifies the radial profile of the gas, making the galaxy more compact for intermediate radii and inducing local changes in the depletion times.

In the isolated galaxy, collapse is triggered in the Toomre unstable part of the disc. In this region the ISM is maximally star forming in the sense that there is no locked-up gas that is not accessible to SF, and the rate is self-regulated by feedback from young stellar populations injecting energy into the system to counterbalance the mid-plane pressure. Since the interaction cannot drastically change the latter, the ISM changes only slightly. On the other hand, the outer regions of the disc, which were stable in the isolated case, are unable to form stars even in the interacting simulation since the encounter ejects most of this gas into the extended atomic tidal tail.

We conclude that SFR and the balance of the gas between the different phases is set by self-regulation in response to stellar feedback, and the effect of the interaction is limited here to inducing changes in the morphology of the galaxy, grouping the already present molecular gas and associated SF into dense spiral arms.

\section*{Acknowledgements}
We thank Mattis Magg, Ondrej Jaura, Eric Pellegrini and Ana Duarte Cabral Peretto for insightful comments and discussions. We further thank R\"udiger Pakmor and Volker Springel for letting us use their code. MCS, RGT, SCOG, and RSK acknowledge support from the Deutsche Forschungsgemeinschaft via the Collaborative Research Centre (SFB 881) ``The Milky Way System'' (subprojects B1, B2, and B8) and the Priority Program SPP 1573 ``Physics of the Interstellar Medium'' (grant numbers KL 1358/18.1, KL 1358/19.2, and GL 668/2-1). RGT also thanks the AMNH and the Kade foundation for its support and hospitality in the early stages of this project. RSK furthermore thanks the European Research Council for funding in the ERC Advanced Grant STARLIGHT (project number 339177). RJS gratefully acknowledges an STFC Ernest Rutherford fellowship (grant ST/N00485X/1) and HPC from the Durham DiRAC supercomputing facility (grants ST/P002293/1, ST/R002371/1, ST/S002502/1, and ST/R000832/1). The authors acknowledge support by the state of Baden-W\"urttemberg through bwHPC and the German Research Foundation (DFG) through grant INST 35/1134-1 FUGG. M-MML acknowledges partial support from US NSF grant AST-1815461, and the hospitality of the Insitut f\"ur Theoretische Astrophysik during writing of this paper. PCC acknowledges support from the Science and Technology Facilities Council (under grant ST/N00706/1). PCC also acknowledges StarFormMapper, a project that has received funding from the European Union's Horizon 2020 Research and Innovation Programme, under grant agreement no. 687528.


\bibliographystyle{mnras}
\bibliography{bibliography}


\appendix

\section{resolution study}
\label{sec:resolutionDependece}

\begin{figure}
	\includegraphics[width=\columnwidth]{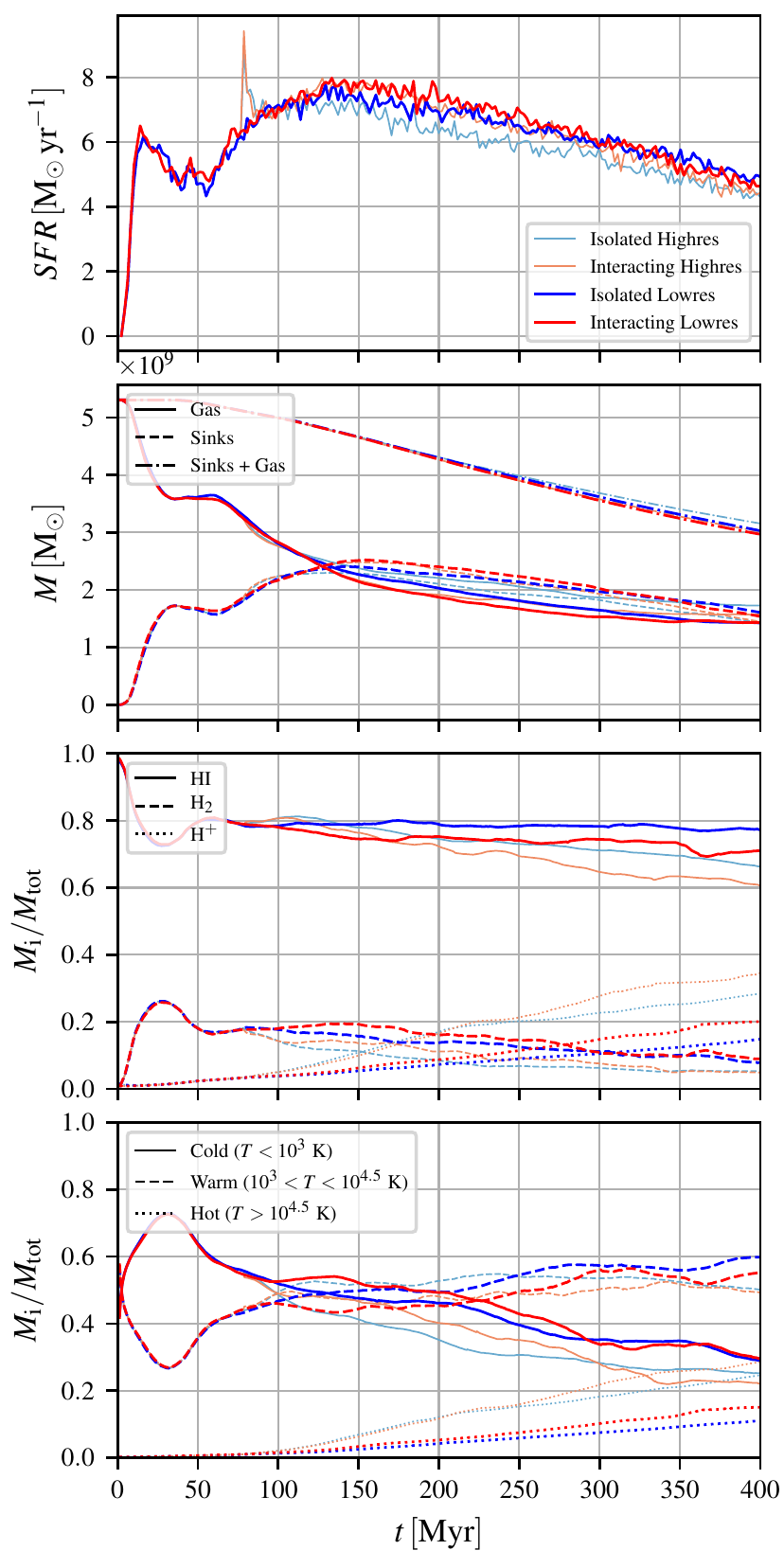}
    \caption{Comparison between the simulations at high and low resolution. Throughout the panels the simulations at high-resolution are shown in {\em orange} (interacting) and {\em light-blue} (isolated) while the isolated simulation at low-resolution is shown in {\em blue} and the interacting low-resolution simulation in {\em red}. In the top panel we compare the SF rates, in the second from top panel we depict the total gas mass {\em (solid line)}, the mass in sink particles {\em (dashed line)} and the total mass of the two combined {\em (dash-dotted line)} as a function of time. The second to last panel shows the H~{\sc i} {\em (solid line)}, H$_2$ {\em (dashed line)} and H$^+$ {\em (dotted line)} fractions as a function of time for the different runs. Finally we plot the cold ($T<10^3$~K), warm ($10^3<T<10^{4.5}$~K) and hot ($T>10^{4.5}$~K) gas fractions in the bottom panel.}
    \label{fig:resolution_study}
\end{figure}

Here we check how robust our results are as a function of the resolution. As described in Section \ref{sec:resolution} we run the simulations at two different resolutions ($1000$~M$_\odot$ and $300$~M$_\odot$ per cell) and we stop the Jeans refinement at one order of magnitude lower densities for the low resolution case. In Fig.\ \ref{fig:resolution_study} we compare the SF and the ISM properties between the two cases. By looking at the SFRs we see that our main finding is rather resolution independent: the difference in stars formed between the isolated and the interacting case stays fairly low. Even the magnitude of the SFR is comparable between the low and the high resolution runs. Given the model used, this is an indication of numerical convergence, although even higher resolution is needed to give a complete interpretation. Unfortunately this is computationally out of reach at the moment as our high resolution runs were already stretching the limits of feasibility (each simulation at $300$~M$_\odot$ per cell required a few months to run with approximately 512 cores).

The second panel shows the mass in sinks and the mass of gas in the simulations as a function of time. Once again, given our physical prescription, the behaviour seems numerically converged. The sinks formed in the low resolution simulation are in general more massive, but less numerous such that their total mass is similar to the high resolution case. 

In the last two panels we look at the ISM. We do not account for the gas mass inside sink particles here since we are interested in comparing the properties of the actual simulated ISM. We note that in this case there is a quantitative difference between the simulations. In particular in the high resolution case more supernovae are actually resolved and the sharper interfaces between hot and cold gas cool less, so higher fractions of the hot ionised phase are reached at the expense of the warm phase. The difference in the molecular phase, on the other hand, is smaller between the two different resolutions. At lower resolutions the substructure of clouds is poorly resolved and GMCs are rather big blobs and as such better shielded from the surrounding interstellar radiation field. This allows for higher molecular fractions in the low resolution case.

Despite these quantitative differences, the impact of the interacting galaxy on the ISM properties stays unchanged and the picture described in the main text holds independently of resolution. The magnitude of the difference in the ISM phases induced by the interaction is the same for the simulations at different resolution and stays contained to less than $10$~\% throughout the computation in both cases.


\bsp	
\label{lastpage}
\end{document}